\documentclass[letters,usenatbib]{mnras}
\usepackage{threeparttable}
\usepackage{newtxtext,newtxmath}

\usepackage[T1]{fontenc}

\DeclareRobustCommand{\VAN}[3]{#2}
\let\VANthebibliography\thebibliography
\def\thebibliography{\DeclareRobustCommand{\VAN}[3]{##3}\VANthebibliography}

\usepackage{graphicx}	
\usepackage{amsmath}	
\newcommand\Nu{NuSTAR }
\newcommand\swift{SWIFT J2127.4+5654 }
\newcommand\NGC{NGC 3227 }
\newcommand\ec{$E_{\rm cut} $ }

\title[The High Energy Cutoff Variations]{Distinct High Energy Cutoff Variation Patterns in Two Seyfert Galaxies}

\author[Jia-Lai Kang et al.]{
Jia-Lai Kang,$^{1,2}$\thanks{ericofk@mail.ustc.edu.cn}
Jun-Xian Wang,$^{1,2}$\thanks{jxw@ustc.edu.cn}
Wen-Yong Kang$^{1,2}$
\\
$^{1}$CAS Key Laboratory for Research in Galaxies and Cosmology, Department of Astronomy, University of Science and Technology of
China, Hefei, Anhui 230026, China\\
$^{2}$School of Astronomy and Space Science, University of Science and Technology of China, Hefei 230026, China\
}

\date{Accepted 2021 January 4. Received 2020 December 22; in original form 2020 September 29}

\pubyear{2020}

\begin{document}
\label{firstpage}
\pagerange{\pageref{firstpage}--\pageref{lastpage}}
\maketitle

\begin{abstract}
Investigating how the cutoff energy $E_{\rm cut}$ varies with X-ray flux and photon index $\Gamma$ in individual AGNs opens a new window to probe the yet unclear coronal physics. So far $E_{\rm cut}$ variations have only been detected in several AGNs but different patterns have been reported. Here we report new detections of $E_{\rm cut}$ variations in two Seyfert galaxies with multiple \Nu exposures. While in \NGC $E_{\rm cut}$ monotonically increases with $\Gamma$, the $E_{\rm cut}$--$\Gamma$ relation exhibits a $\Lambda$ shape in SWIFT J2127.4+5654 ($E_{\rm cut}$ increasing with $\Gamma$ at $\Gamma$ $\lesssim$ 2.05, but reversely decreasing at $\Gamma$ $\gtrsim$ 2.05), indicating more than a single underlying mechanism is involved. Meanwhile both galaxies show softer spectra while they brighten in X-ray, a common phenomenon in Seyfert galaxies. Plotting all 7 AGNs with $E_{\rm cut}$ variations ever reported with \Nu observations in the $E_{\rm cut}$--$\Gamma$ diagram, we find they could be unified with the $\Lambda$ pattern. Although the sample is small and \swift is the only source with $\Gamma$ varying across the break point thus the only one exhibiting the complete $\Lambda$ pattern in a single source, the discoveries shed new light on the coronal physics in AGNs. Possible underlying physical mechanisms are discussed.
\end{abstract}


\section{Introduction}\label{sec:intro}
\par In the standard disc-corona paradigm the hard X-ray emission of active galactic nuclei (AGNs) is produced in a hot and compact region, the so called corona \citep[e.g.][]{Haardt_1991,Haardt_1994}, through the inverse Compton scattering of the seed photons from the accretion disk. This process could produce the observed power law continuum with a high-energy cutoff. 
Such a cutoff has been detected in a number of AGNs \citep{Zdziarski_2000, Molina_2013}, particularly with the high-quality hard X-ray spectra of NuSTAR \citep[e.g.,][]{Matt_2015_highE,Tortosa_2018, Molina_2019,Kang_2020}, providing key constraints on the yet unclear coronal physics \citep[e.g.][]{Fabian_2015}.

\par The Nuclear Spectroscopic Telescope Array \citep[NuSTAR][]{Harrison_2013} also enables the detection of $E_{\rm cut}$ (or coronal temperature $T_e$) variations in individual AGNs with multiple exposures, including 3C 382 \citep{Ballantyne_2014}, NGC 5548 \citep{Ursini_2015}, Mrk 335 \citep{Keek_2016}, NGC 4593 \citep{Ursini_2016, Middei_2019}, MCG--5--23--16 \citep{Zoghbi_2017}, 
and 4C 74.26 \citep{Zhangjx2018}.  \citet{Zhangjx2018} also revisited the first five sources mentioned above using the spectra ratio technique they developed. They confirmed the claimed $E_{\rm cut}$ variations in 3 of them, but disproved those in NGC 4593 and MCG--5--23--16. 
Despite the limited number of sources,  these studies have opened a new window to probe the coronal physics. 
Remarkably, \citet{Zhangjx2018} found that all the 4 AGNs with \ec variations confirmed tend to have larger $E_{\rm cut}$ (thus hotter corona) when they brighten and soften in X-ray.
In other words, they show a ``hotter-when-brighter'' behavior along with the conventional ``softer-when-brighter'' pattern \citep[e.g.,][]{Markowitz_2003,Sobolewska_2009}. Possible underlying mechanisms have been discussed in \citet{Zhangjx2018} and \citet{Wu_2020}, including geometrical changes of the corona and pair production.
Do Seyfert galaxies universally follow this ``hotter-when-softer/brighter'' pattern? A possible counter-example is the narrow line Seyfert 1 galaxy (NLS1) Ark 564, for which \citet{Barua_2020} found cooler corona (though statistically marginal) during the softer and brighter phases within a 200 ks \Nu observation. 

\par In this letter we report new detections of \ec variations in two Seyfert galaxies, \NGC and SWIFT J2127.4+5654, and a $\Lambda$ shaped $E_{\rm cut}$--$\Gamma$ relation for the first time. \S \ref{sec:obs} presents the NuSTAR observations and data reduction. In \S \ref{sec:fitting} we describe the spectral models and deliver the fitting results. In \S \ref{sec:discussion} we discuss the spectral variabilities and the underlying mechanisms for AGNs with reported \ec variations. 

\begin{figure*}
\includegraphics[width=7in]{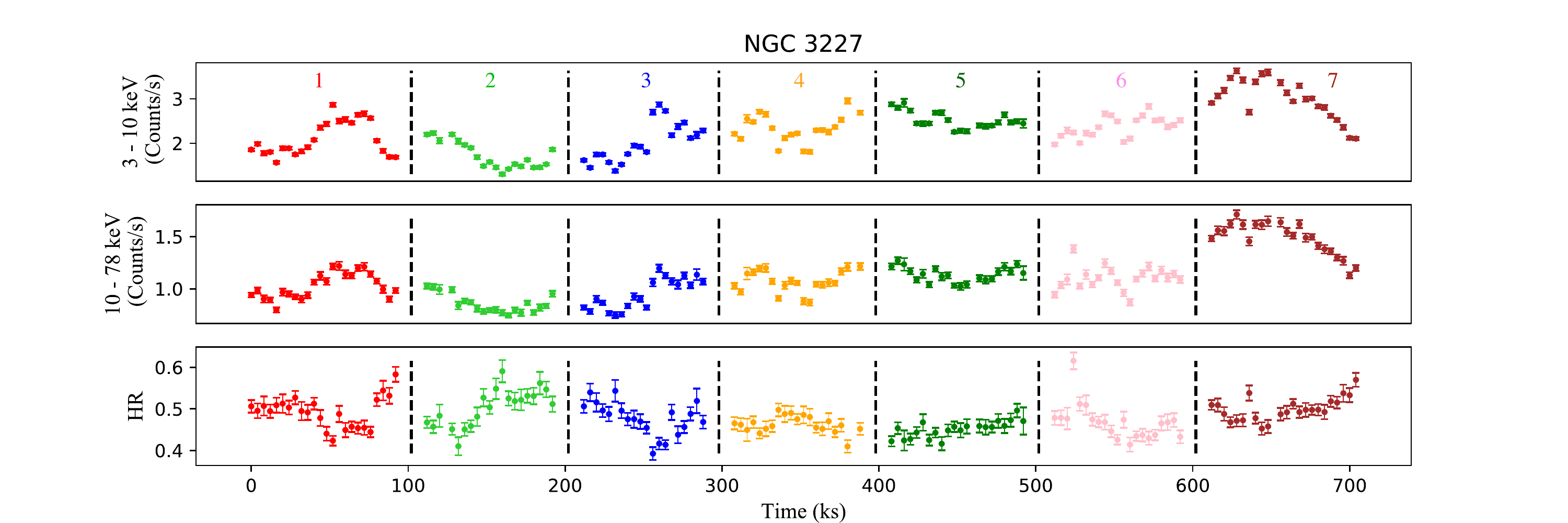}\\
\includegraphics[width=7in]{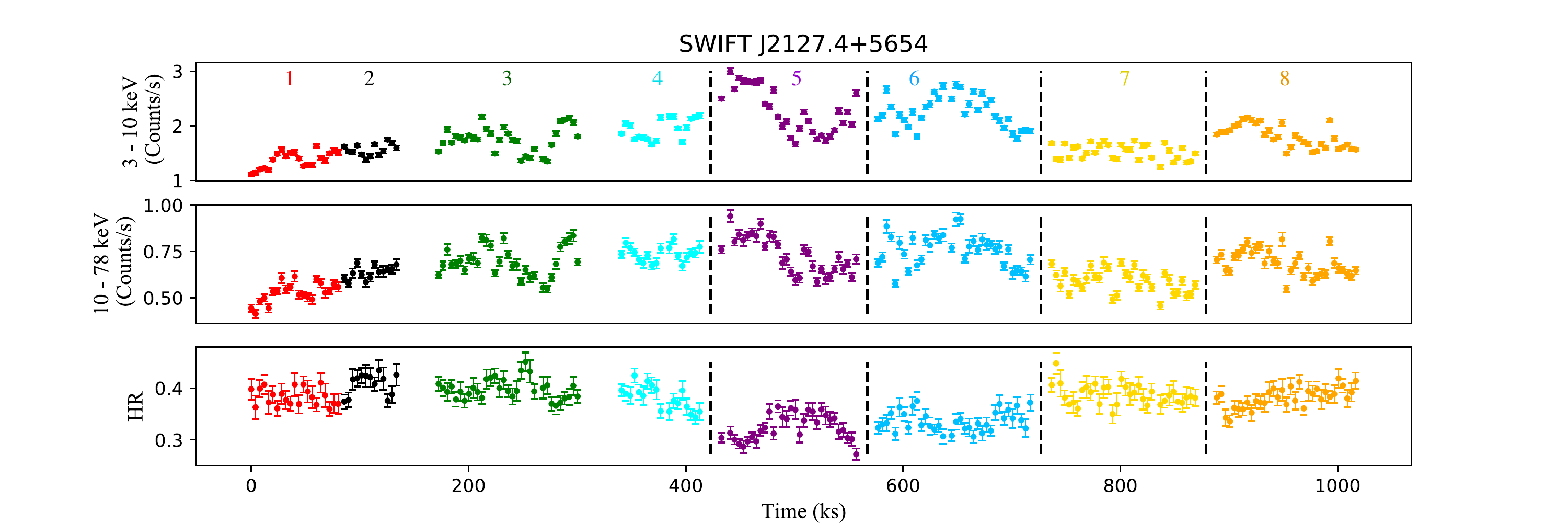}\\
\caption{NuSTAR light curves (with a bin size of 4 ks) in 3 -- 10 keV, 10 -- 78 keV bands and the 10 -- 78 keV / 3 -- 10 keV hardness ratios. 
The observations, sorted and numbered by observation date (see Tab. \ref{tab:results}), are color coded. The vertical dashed lines mark the positions where the x-axis is discontinuous because of long intervals between exposures. 
 }
\label{fig:all_LC}
\end{figure*}

\begin{figure*}
\includegraphics[width=3.3in]{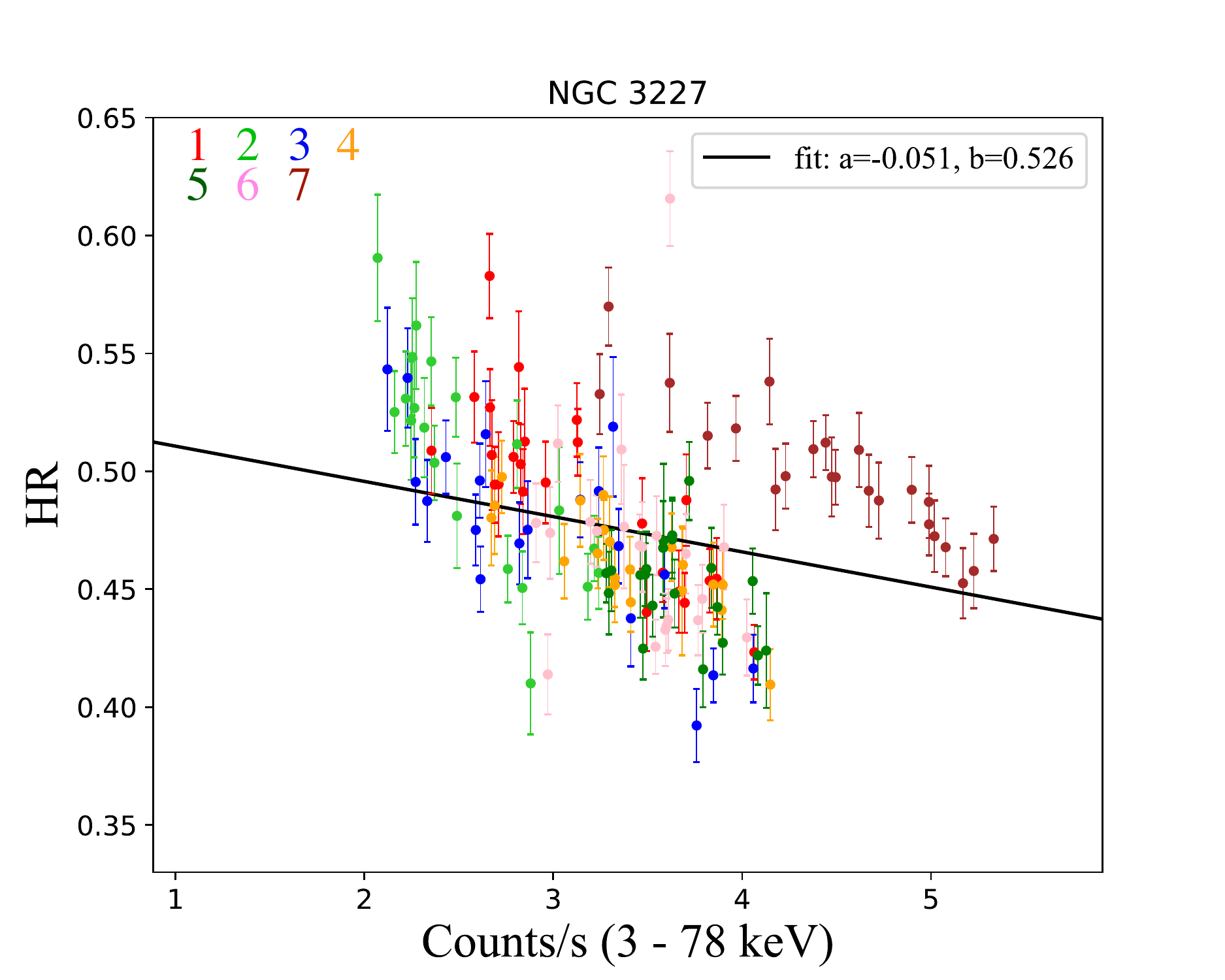}
\includegraphics[width=3.3in]{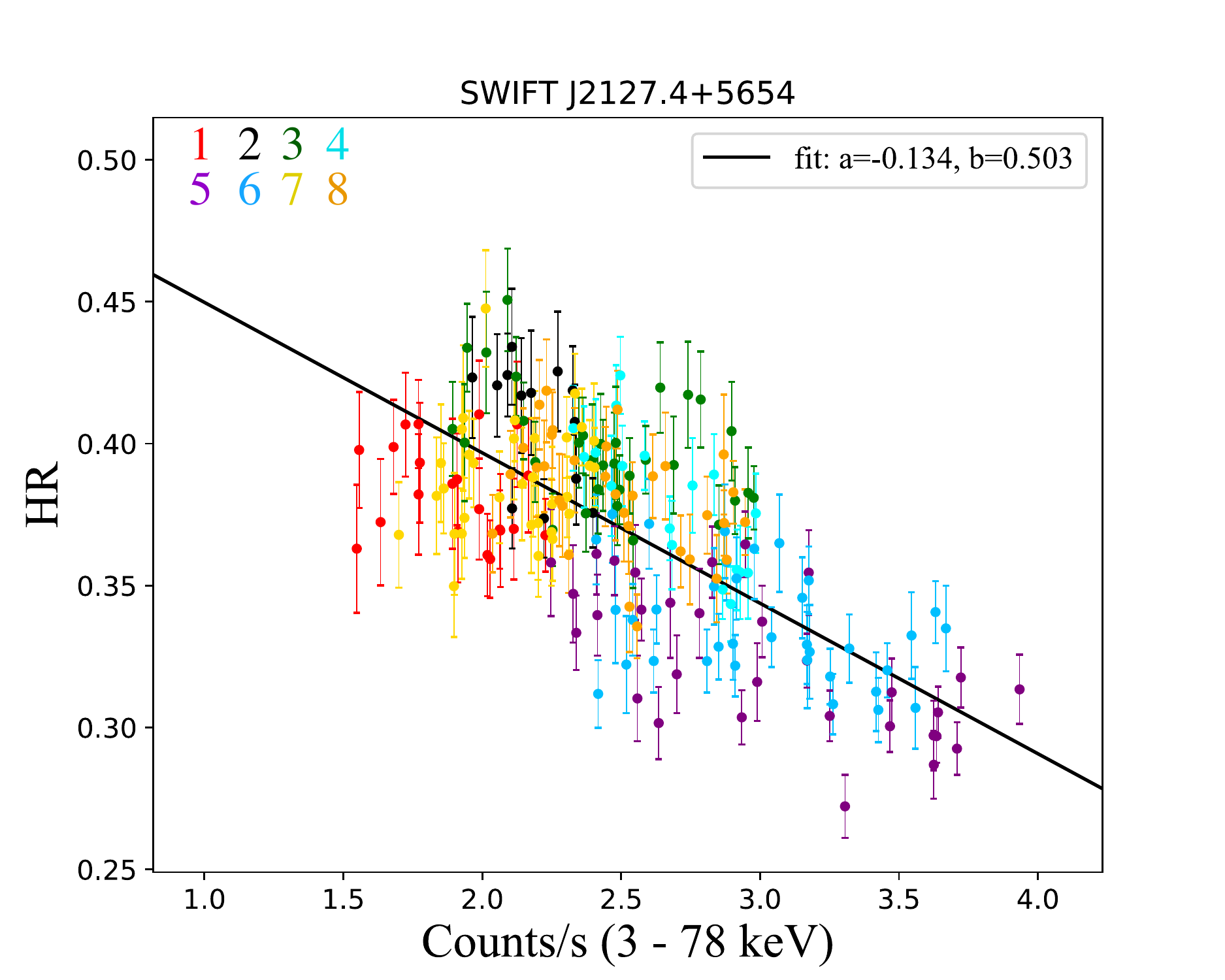}\\
\caption{The 10 -- 78 keV / 3 -- 10 keV hardness ratio versus 3 -- 78 keV count rate. The black lines show simple linear fits to the data points.
The observations are color coded as in Fig. \ref{fig:all_LC}. 
 }
\label{fig:all_HR}
\end{figure*}

\begin{table} \scriptsize
	\centering
	\caption{NuSTAR Observation Logs.}\label{tab:details}
	\begin{tabular}{ccccc}
\hline
 Source & ID and No. & Obs. time& Exposure &  Flux$_{\rm 3-78 keV}$\\
&  &  & (ks) & $(10^{-10} erg/cm^2/s)$\\
\hline
NGC 3227 & 60202002002 (1) & 2016--11--09 & 49.8 & 1.26 \\
 & 60202002004 (2) & 2016--11--25 & 42.5 & 1.05 \\
 & 60202002006 (3) & 2016--11--29 & 39.7 & 1.16 \\
 & 60202002008 (4) & 2016--12--01 & 41.8 & 1.37 \\
 & 60202002010 (5) & 2016--12--05 & 40.9 & 1.38 \\
 & 60202002012 (6) & 2016--12--09 & 39.3 & 1.35 \\
 & 60202002014 (7) & 2017--01--21 & 47.6 & 1.8 \\
SWIFT J2127.4+5654 & 60001110002 (1) & 2012--11--04 & 49.2 & 0.58\\
 & 60001110003 (2)& 2012--11--05 & 28.8 & 0.71\\
 & 60001110005 (3)& 2012--11--06 & 74.6 & 0.77\\
 & 60001110007 (4)& 2012--11--08 & 42.1 & 0.82\\
 & 60402008004 (5)& 2018--07--16 & 71.6 & 0.83\\
 & 60402008006 (6)& 2018--07--30 & 72.1 & 0.85\\
 & 60402008008 (7)& 2018--09--14 & 72.9 & 0.66\\
 & 60402008010 (8)& 2018--12--30 & 74.2 & 0.76\\
\hline
\end{tabular}
\begin{tablenotes}
\item{$\star$}: {For convenience, the observations for each source were sorted and numbered (in parenthesis) by time.}
\end{tablenotes}
\end{table}

\begin{table*}\scriptsize
	\centering
	\caption{Spectral Fitting Results.}\label{tab:results}
	\renewcommand\tabcolsep{3.0pt}
	\begin{tabular}{ccccccccccccccc} 
\hline
 ID and No. &  N$_{\rm H}$& $\Gamma$& $R$& EW&  $E_{\rm pexrav}$&  $\chi^2_\nu$ (pexrav)&  $E_{\rm pexriv}$ &  $\chi^2_\nu$ (pexriv) &  $E_{\rm relxill}$ &  $\chi^2_\nu$ (relxill)&  $kT_{\rm xillverCp}$ &  $\chi^2_\nu$ (xillverCp) &  $kT_{\rm relxillCp}$ &  $\chi^2_\nu$ (relxillCp) \\
  &   ($10^{22} cm^{-2}$)  &  &  &(eV)& (keV)   &  & (keV) &  & (keV)&  & (keV) &   & (keV) & \\
\hline
\multicolumn{15}{c}{\NGC}\\
\hline
 60202002002 (1) &$1.4^{+0.5}_{-0.5}$ & $1.75^{+0.06}_{-0.06}$ & $0.94^{+0.23}_{-0.2}$ & $100^{+16}_{-16}$ & $185^{+115}_{-53}$  & 1.05 &$248^{+254}_{-87}$ &  1.05 &$128^{+43}_{-27}$ &  1.09 &$24^{+5}_{-3}$ &  1.1 & $24^{+5}_{-3}$ &  1.11 \\
 60202002004 (2) &$0.5^{+0.8}_{-0.5}$ & $1.63^{+0.09}_{-0.08}$ & $0.73^{+0.26}_{-0.22}$ & $211^{+49}_{-43} \star $ & $113^{+54}_{-28}$  & 0.96 &$126^{+68}_{-34}$ &  0.95 &$96^{+35}_{-17}$ &  1.00 &$22^{+4}_{-4}$ &  1.02 & $18^{+6}_{-1}$ &  1.03 \\
 60202002006 (3) &$1.5^{+0.6}_{-0.6}$ & $1.84^{+0.07}_{-0.07}$ & $0.88^{+0.26}_{-0.22}$ & $97^{+18}_{-19}$ & $ > 247$ &  1.16 &$ > 280$ &  1.15 &$236^{+300}_{-86}$ &  1.15 &$ > 35$ &  1.17 & $ > 74$ &  1.19 \\
 60202002008 (4) &$1.1^{+0.4}_{-0.4}$ & $1.88^{+0.05}_{-0.05}$ & $0.88^{+0.22}_{-0.2}$ & $88^{+16}_{-16}$ & $ > 780$ &  1.02 &$ > 1000$ &  1.03 &$ > 350$ &  1.04 &$ > 43$ &  1.03 & $ > 65$ &  1.06 \\
 60202002010 (5) &$0.8^{+0.5}_{-0.5}$ & $1.88^{+0.06}_{-0.06}$ & $1.04^{+0.26}_{-0.22}$ & $54^{+16}_{-16}$ & $ > 218$ &  0.99 &$ > 260$ &  0.97 &$ > 233$ &  0.98 &$ > 35$ &  1.02 & $ > 42$ &  0.99 \\
 60202002012 (6) &$2.0^{+0.3}_{-0.5}$ & $1.93^{+0.03}_{-0.06}$ & $1.09^{+0.29}_{-0.24}$ & $74^{+17}_{-17}$ & $ > 247$ &  0.93 &$ > 680$ &  0.92 &$ > 210$ &  0.94 &$ > 31$ &  1.00 & $ > 33$ &  0.95 \\
 60202002014 (7) &$4.1^{+0.4}_{-0.4}$ & $1.9^{+0.05}_{-0.05}$ & $1.21^{+0.22}_{-0.19}$ & $52^{+13}_{-13}$ & $355^{+473}_{-133}$  & 1.01 &$ > 350$ &  1.01 &$172^{+74}_{-39}$ &  1.01 &$ > 36$ &  1.04 & $ > 49$ &  1.04 \\
\hline
\multicolumn{15}{c}{\swift}\\
\hline
 60001110002 (1) &$ < 0.0$ & $1.88^{+0.05}_{-0.05}$ & $1.54^{+0.43}_{-0.36}$ & $135^{+39}_{-35} \star $ & $63^{+15}_{-10}$ &   1.01 &$71^{+18}_{-12}$ &  1.01 &$110^{+41}_{-26}$ &  1.06 &$19^{+2}_{-2}$ &  1.07 &$21^{+9}_{-2}$ &  1.04 \\
 60001110003 (2) &$0.9^{+0.8}_{-0.8}$ & $2.0^{+0.1}_{-0.1}$ & $1.58^{+0.54}_{-0.43}$ & $41^{+24}_{-24}$ & $109^{+80}_{-34}$ &   0.96 &$135^{+146}_{-48}$ &  0.97 &$181^{+115}_{-55}$ &  0.98 &$35^{+26}_{-13}$ &  1.03 &$38^{+33}_{-14}$ &  0.98 \\
 60001110005 (3) &$1.3^{+0.4}_{-0.4}$ & $2.04^{+0.06}_{-0.06}$ & $1.95^{+0.35}_{-0.31}$ & $59^{+14}_{-14}$ & $92^{+26}_{-17}$ &   1.01 &$117^{+45}_{-26}$ &  1.02 &$121^{+24}_{-23}$ &  1.04 &$21^{+2}_{-2}$ &  1.13 &$26^{+7}_{-5}$ &  1.03 \\
 60001110007 (4) &$0.6^{+0.6}_{-0.6}$ & $1.98^{+0.08}_{-0.08}$ & $1.65^{+0.41}_{-0.35}$ & $44^{+17}_{-18}$ & $67^{+18}_{-12}$ &   0.96 &$76^{+25}_{-15}$ &  0.96 &$108^{+32}_{-18}$ &  1.02 &$18^{+1}_{-2}$ &  1.03 &$21^{+7}_{-2}$ &  1.00 \\
 60402008004 (5) &$0.9^{+0.4}_{-0.4}$ & $2.17^{+0.06}_{-0.06}$ & $1.83^{+0.33}_{-0.29}$ & $60^{+12}_{-12}$ & $63^{+12}_{-9}$ &   1.09 &$69^{+15}_{-10}$ &  1.07 &$93^{+16}_{-15}$ &  1.12 &$16^{+1}_{-1}$ &  1.18 &$24^{+6}_{-4}$ &  1.10 \\
 60402008006 (6) &$0.8^{+0.5}_{-0.5}$ & $2.11^{+0.06}_{-0.06}$ & $1.64^{+0.32}_{-0.28}$ & $85^{+34}_{-27} \star $ & $72^{+17}_{-11}$ &   0.93 &$85^{+23}_{-15}$ &  0.94 &$100^{+23}_{-12}$ &  0.96 &$18^{+1}_{-1}$ &  1.03 &$28^{+9}_{-6}$ &  0.95 \\
 60402008008 (7) &$0.4^{+0.6}_{-0.4}$ & $1.98^{+0.08}_{-0.07}$ & $1.81^{+0.41}_{-0.35}$ & $115^{+41}_{-34} \star $ & $81^{+24}_{-15}$ &   0.95 &$97^{+36}_{-21}$ &  0.95 &$143^{+36}_{-26}$ &  0.98 &$23^{+4}_{-3}$ &  1.07 &$33^{+10}_{-8}$ &  0.97 \\
 60402008010 (8) &$1.3^{+0.4}_{-0.4}$ & $2.12^{+0.06}_{-0.06}$ & $2.27^{+0.39}_{-0.34}$ & $58^{+14}_{-14}$ & $90^{+25}_{-16}$ &   1.03 &$113^{+42}_{-25}$ &  1.05 &$123^{+26}_{-22}$ &  1.09 &$18^{+1}_{-1}$ &  1.19 &$28^{+8}_{-5}$ &  1.07 \\
 \hline
\end{tabular}
\begin{tablenotes}
\item{$\star$}: {In some observations a broad Fe K$\alpha$ line is statistically needed, with the corresponding EW marked with $\star$. NGC 3227: 60202002004, $\sigma$=$0.24_{-0.10}^{+0.11}$ keV. SWIFT J2127.4+5654: 60001110002, $\sigma$=$0.28_{-0.12}^{+0.14}$ keV; 60402008006, $\sigma$=$0.29_{-0.15}^{+0.20}$ keV; 60402008008, $\sigma$=$0.28_{-0.13}^{+0.16}$ keV.}
\end{tablenotes}
\end{table*}

\section{Observations and Data Reduction} \label{sec:obs}

 \NGC is a radio-quiet Seyfert 1.5 galaxy \citep{Classification} at z = 0.0039\footnote{The radio types and the redshifts are from NED: http://ned.ipac.caltech.edu}. In Tab. \ref{tab:details} we list the seven archival \Nu observations of NGC 3227. NGC 3227 shows highly variable X-ray emission and its absorption feature have been extensively investigated in literature. As an example, based on the \Nu and XMM-Newton observations, \citet{Turner_2018} found a rapid occultation event in NGC 3227 between two exposures of \Nu (60202002010 and 60202002012), with several absorber zones involved. However, as further shown in \S\ref{sec:fitting}, our study on the high-energy cutoff in this work is barely influenced by these complex absorbers. 

\par Meanwhile, SWIFT J2127.4+5654, a radio-quiet NLS1 \citep{Malizia_2008} at z = 0.0144, has been observed by NuSTAR in two campaigns, with four exposures in 2012 and five in 2018. We dropped the observation 60402008002 (not listed in Tab. \ref{tab:details}) of SWIFT J2127.4+5654, which has an issue flag = 1, indicating possible contamination from solar activity or other unexpected issues. \citet{Marinucci_2014} performed a joint spectral fitting of the four 2012 observations with (quasi-)simultaneous XMM-Newton data and an average \ec = $108^{+11}_{-10}$ keV was reported. Besides, using XMM-Newton data only, \citet{Sanfrutos_2013} reported a partial covering absorber in \swift with $N_{\rm H} = 2 \times 10^{22} cm^{-2}$ and a covering fraction $\sim$ 0.43, whereas we find it has negligible effect in \Nu spectra.  

\par \Nu data reduction is performed using the NuSTAR Data Analysis Software (NuSTARDAS) within the HEAsoft package (version 6.26). With CALDB version 20190513, the calibrated and cleaned event files are produced with \textit{nupipeline}. We first extract the light curves using \textit{nuproducts}, adopting a circular source region with a radius of 60\arcsec\ centered on each source, and an annulus from 120\arcsec\ to 200\arcsec\ for background extraction. The light curves from FPMA and FPMB, with the livetime, PSF/EXPOSURE and vignetting corrections applied, are then combined using \textit{lcmath}. The 3 -- 10 keV and 10 -- 78 keV light curves, along with the 10 -- 78 keV/ 3 -- 10 keV hardness ratios, are plotted in Fig. \ref{fig:all_LC}. Both sources show clear variations in flux (count rate) and spectra shape (hardness ratio), not only between but also within the individual exposures. In the plot of hardness ratio (HR) versus count rate (Fig. \ref{fig:all_HR}) both sources clearly exhibit the ``softer-when-brighter'' pattern. However different individual exposures appear to follow different ``softer-when-brighter'' tracks. For instance, Obs. ID 60202002014 (No. 7) of NGC 3227 clearly deviate from other exposures in Fig. \ref{fig:all_HR}. See also exposures No. 1 and No. 2, No. 5 and No. 6 of SWIFT J2127.4+5654. 
Such variations could be due to the physical or structural changes in the corona which may lead to different ``softer-when-brighter" tracks \citep[e.g., ][]{Sarma2015}, stopping us from
merging data from different exposures according to the count rate or hardness ratio. In this work we focus on the analyses of spectra integrated over individual exposures and investigate the $E_{\rm cut}$ variations between these exposures. 
Though rapid hardness ratio variations within individual exposures are seen, due to limited photon counts we are yet unable to explore more rapid $E_{\rm cut}$ variations in two sources.

\par Source spectra are extracted from the same circular regions as the light curves, using the \textit{nuproducts}. As for background extraction, we use NUSKYBGD developed by \citet{Wik_2014} to handle the spatially variable background of \Nu observations \citep[see also][for an example]{Kang_2020}. NuSTAR observations generally do not suffer from pile-up\footnote{https://heasarc.gsfc.nasa.gov/docs/nustar/nustar\_faq.html\#pileup}. Meanwhile, we find the low-energy effective area issue for FPMA \citep{Madsen_2020} to be insignificant in all the observations and no low energy excess in FPMA spectra is found (see Fig. \ref{fig:all_spectra}). As a final step, the source spectra are rebinned to achieve a minimum of 50 counts bin$^{-1}$ using \textit{grppha}.

\par We notice there are (quasi-)simultaneous XMM-Newton observations for both sources (six of \NGC and three of SWIFT J2127.4+5654). However, we find slightly different photon indices between most XMM-Newton and \Nu exposures for both sources. Such discrepancy, likely due to the inter-instrument calibration, is widely found in literature \citep[e.g.][]{gamma_xmm_1, gamma_xmm_3, gamma_xmm_2}. Additionally, the fact that XMM-Newton and \Nu exposures are not perfectly simultaneous may also have played a role due to the intrinsic spectral variation. However requiring perfect simultaneity would yield significant loss of the valuable \Nu exposure time. Since $E_{\rm cut}$ measurement is sensitive to the photon index \citep[e.g.,][]{Molina_2019, Kang_2020} and not all \Nu exposures have corresponding XMM-Newton observations, in this work we do not include those XMM-Newton exposures but provide uniform spectral fitting to \Nu spectra alone. 

\section{Spectral fitting} \label{sec:fitting}

\begin{figure*}
\includegraphics[width=2.3in]{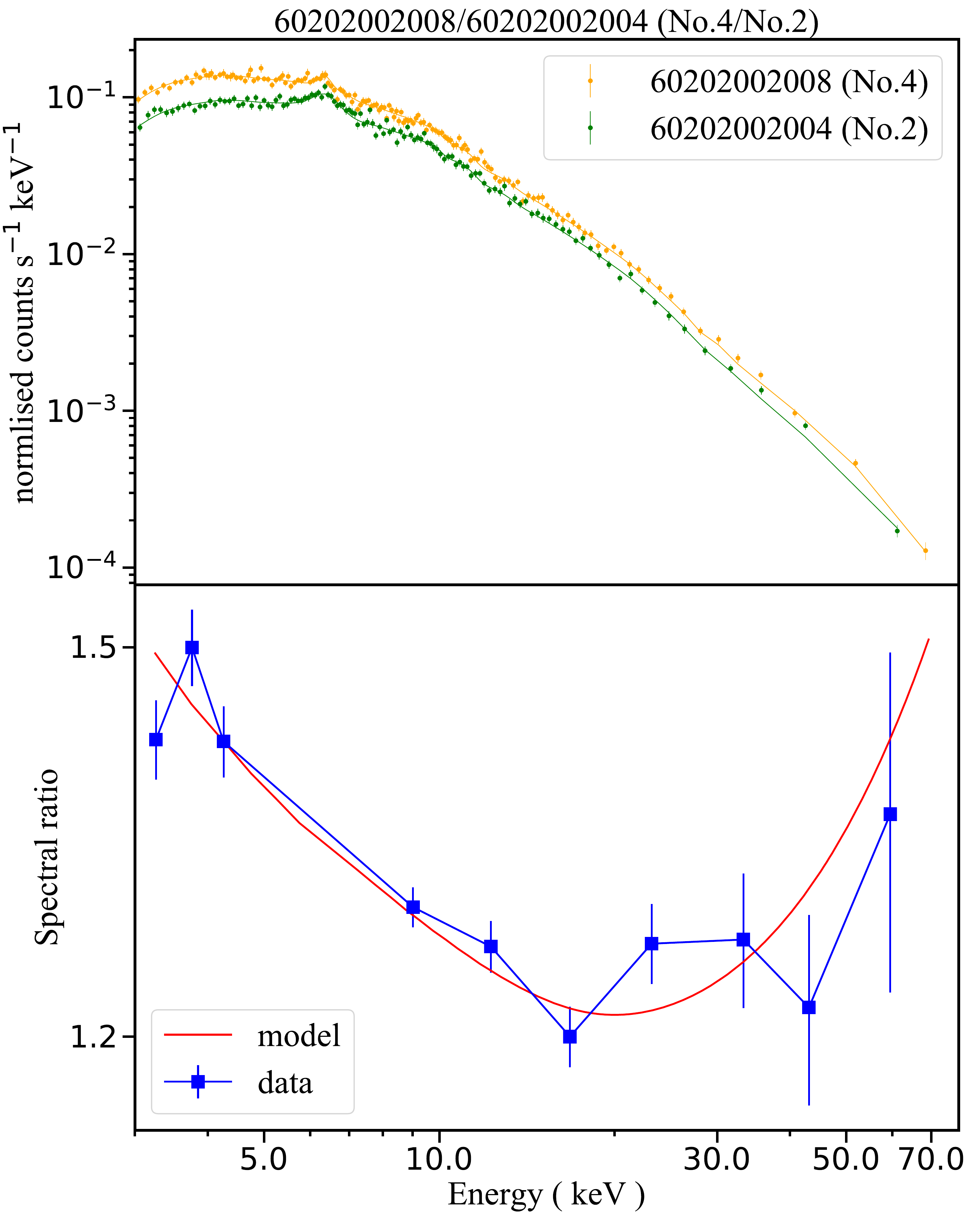}
\includegraphics[width=2.3in]{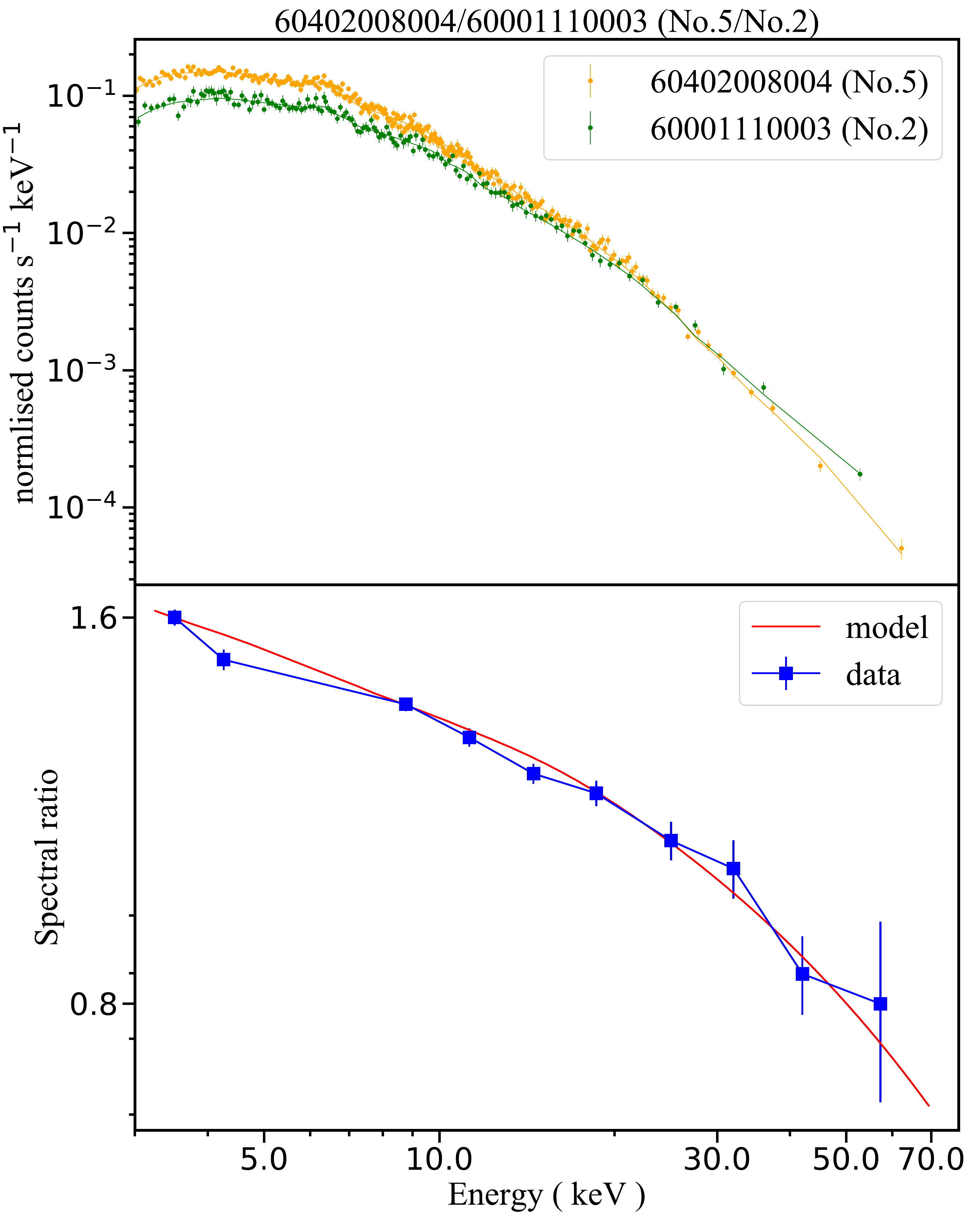}
\includegraphics[width=2.3in]{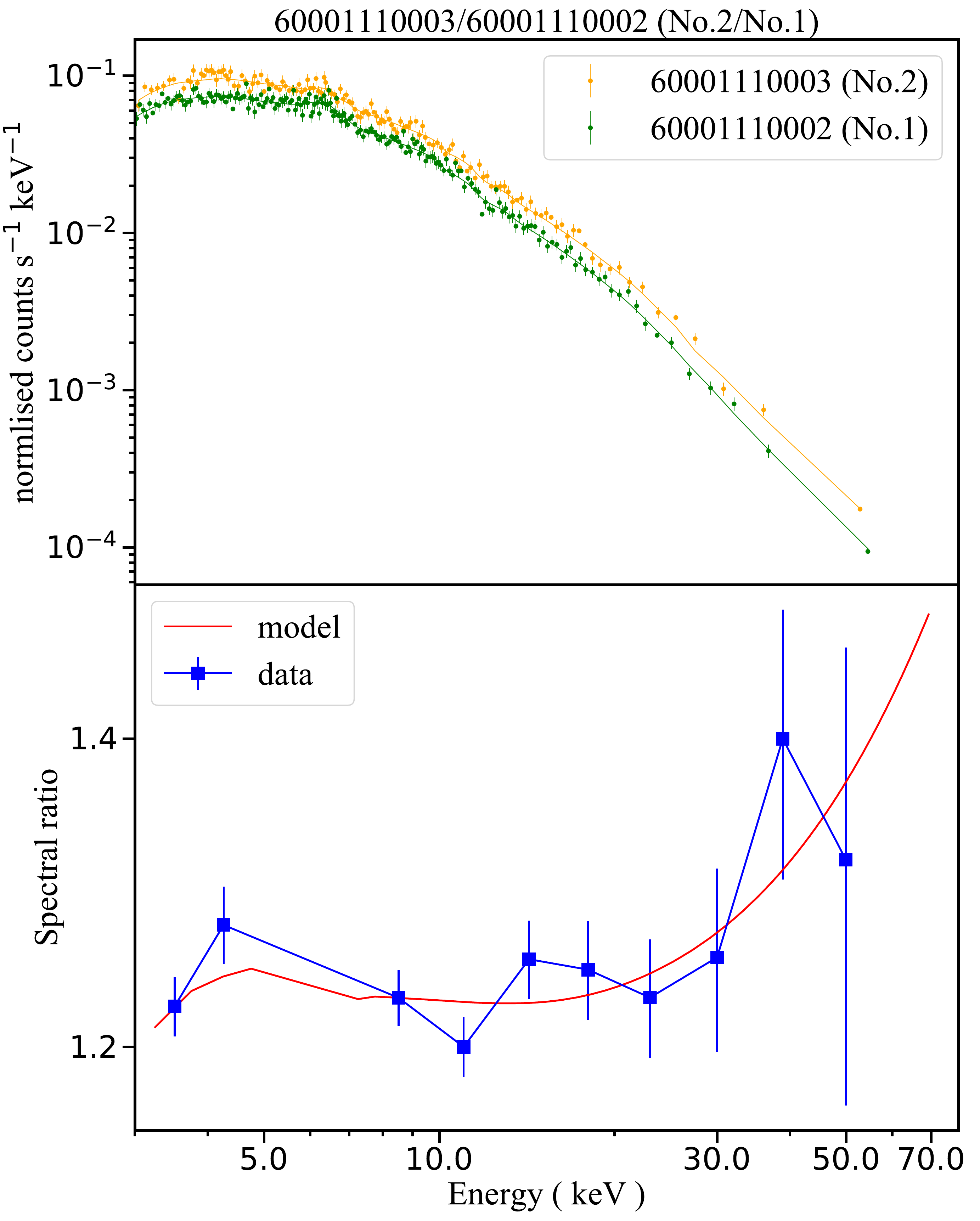}
\caption{The rebinned FPMA spectra and spectral ratio (always the ratio of a brighter spectrum to a fainter one) of \NGC(the left panel) and \swift(the middle and right panels). Following \citet{Zhangjx2018}, we adopt a single power law to fit each spectrum and derive the corresponding unfolded spectra for the calculation of spectra ratio. As approved in \citet{Zhangjx2018}, the spectra ratio plot is insensitive to the adopted spectral model, except for in the spectral range of  Fe K$\alpha$ line (which is dropped from the plot). For NGC 3227 we plot the ratio of the two observations showing the most prominent $E_{\rm cut}$ variation. The upward curvature in the lower left panel (deviation from a straight line at high energies) demonstrates clear ``hotter-when-brighter'' pattern in NGC 3227. For \swift we plot the ratios of two observation pairs, showing both ``cooler-when-brighter'' (the downward high energy curvature in the lower middle panel) and ``hotter-when-brighter'' (the lower right panel) patterns. 
In the lower panels , the ratios of best-fit models presented in this work are over-plotted, illustrating the $E_{\rm cut}$ variations have been properly accounted in the spectral models. 
 }
\label{fig:ratio}
\end{figure*}

\par \citet{Zhangjx2018} developed the spectral ratio technique, analogous to the difference-imaging technique in astronomy, to assist the study of $E_{\rm cut}$ variations. 
Briefly, if the \ec is invariable within two observations, the ratio of two spectra (primarily exponentially cutoff power law) is supposed to be a straight line in log-log space in the spectra ratio plot. So we can notice potential \ec variations directly by looking for deviations from the straight line at the high energy end \citep[see][for details]{Zhangjx2018}. 
We applied this technique to AGNs with multiple archival \Nu exposures and took notice of \NGC and SWIFT J2127.4+5654, of which the spectral ratios indicate clear and prominent \ec variations as shown in Fig. \ref{fig:ratio}. In this section, we perform spectral fitting to quantify the \ec variations in both sources.    
 
\par Spectral fitting is performed in the 3--78 keV band within XSPEC \citep{Arnaud_1996}, using ${\chi}^2$ statistics and the relative element abundances given by \citet{ANDERS1989197}. All errors along with the upper/lower limits reported throughout the letter are calculated using ${\Delta\chi}^2$ = 2.71 criterion (90\% confidence range). For each observation, the spectra obtained by the two NuSTAR modules (FPMA and FPMB) are fitted simultaneously, with a cross-normalization difference typically less than 5\% \citep{Madsen_2015}.

\par We employ \textit{pexrav} \citep[][]{Magdziarz_1995} to describe the exponentially cutoff power law plus the reflection component. For simplicity, the solar element abundance for the reflector and an inclination of $cosi = 0.45$ are adopted (as default of this model)\footnote{These two parameters are poorly constrained with NuSTAR spectra if allowed free to vary, thus are commonly fixed at the default values in many studies on NuSTAR spectra \citep[e.g.][]{Zhangjx2018,Molina_2019,PW2020}.Through joint fitting XMM-Newton and NuSTAR spectra of SWIFT J2127.4+5654, \cite{Marinucci_2014} reported an iron abundance A$_{\rm Fe}$ of 0.71 and an inclination angle of 49$^{\circ}$, slightly different from the default values we adopted. 
Generally, larger A$_{\rm Fe}$ (inclination angle) would yield  slightly higher (higher) reflection fraction R, smaller (larger) photon index $\Gamma$ and lower (higher) cutoff energy $E_{\rm cut}$.
However adopting different values of A$_{\rm Fe}$ or inclination angle would not alter the main results of this work.
}. We let the photon index $\Gamma$, $E_{\rm cut}$ and the reflection fraction $R$ free to vary. In addition, \textit{zphabs} is used to model the intrinsic absorption, with the Galactic absorption ignored due to its negligible impact on \Nu spectra. 
Replacing the $zphabs$ component with $zxipcf$ to account for the partial covering reported in both sources \citep{Turner_2018, Sanfrutos_2013} does not affect our \ec measurements, as such 
absorbers have negligible influence on \Nu spectra. 
 
\par As for the Fe K emission lines, several lines have been previously found in both sources with XMM-Newton spectra \citep[e.g.][]{Markowitz_2009,Marinucci_2014}, whereas some of them are insignificant in \Nu spectra, likely due to the limited spectral resolution ($\sim$ 0.4 keV at 6 keV). We use a statistical standard to model possible lines as follows. We add a \textit{zgauss} component at 6.4 keV (rest frame) with its width fixed at 19 eV \citep[the mean Fe K$\alpha$ line width in AGNs measured with Chandra HETG,][]{Shu_2010} to describe a neutral and narrow Fe K$\alpha$ line. Since the Fe K$\alpha$ line could be relativistically broadened, we allow the width to vary freely. If a free width can significantly improve the fit ($\Delta {\chi}^2 > 5$), the corresponding fitting results are adopted (see Tab. \ref{tab:results}). Furthermore, we find no other potential lines in all the observations with this criterion. To sum up, our final model in XSPEC form is $constant \times zphabs \times (pexrav + zgauss)$. The spectra and the best-fit models are shown in Fig. \ref{fig:all_spectra}. 

\par Besides, we adopt different models to check the results. Firstly we replace the $pexrav$ component with $pexriv$ \citep{Magdziarz_1995} to take account of potential ionized reflection. As shown in Tab. \ref{tab:results}, the results of the two models ($pexrav$ and $pexriv$) are very close. Furthermore, we employ \textit{relxill} \citep{Dauser_2010, Garc_2014} which models the spectra with a cutoff power law and relativistic ionized reflection from the accretion disc. For \swift we fix its spin $a = 0.58$ referring to \citet{Marinucci_2014}, while for \NGC $a$ is tied among the observations during fitting, with $log\xi$ and $A_{\rm Fe}$ tied among observations for both sources. In this way, we derive $a=-0.36_{-0.22}^{+0.31}$, $log\xi=2.65_{-0.16}^{+0.07}$ and $A_{\rm Fe}=2.78_{-0.58}^{+0.61}$ for NGC 3227, and $log\xi=2.80_{-0.04}^{+0.05}$, $A_{\rm Fe}=1.20_{-0.25}^{+0.42}$ for SWIFT J2127.4+5654. We let photon index $\Gamma$, reflection fraction and $E_{\rm cut}$ vary freely, and set other parameters including inclination and disk radii as default. As shown in Tab. \ref{tab:results}, while the \ec results from two models are generally consistent within statistical uncertainties, the \textit{relxill} model yields systematically smaller \ec for \NGC and larger \ec for \swift compared with the $pexrav$ model. Nevertheless, the yielded \ec variation trends from two models, as the focus of this work, are similar. Moreover, we adopt comptonized models $xillverCp$ and $relxillCp$ \citep[][for the normal and relativistic reflection components respectively]{Dauser_2010, Garc_2014} to directly measure the coronal temperature $T_{\rm e}$. The derived coronal temperatures $kT_{\rm relxillCp}$ and $kT_{\rm xillverCp}$ (see Tab. \ref{tab:results}) for SWIFT J2127.4+5654 are generally consistent with 1/3 of the measured $E_{\rm pexrav}$ \citep[e.g.][]{Petrucci_2001}, but systematically smaller than 1/3 of the best-fit $E_{\rm pexrav}$ for NGC 3227. 

\par From Tab. \ref{tab:results} we can see that $pexrav$ and $pexriv$ yield similarly smaller reduced $\chi^2$ compared with other models, while $pexriv$ has one more free parameter than $pexrav$. 
As $pexrav$  is widely adopted in literature to measure \ec in literature \citep[e.g.][]{Zhangjx2018, Molina_2019, PW2020}, to directly compare with those studies, hereafter we simply adopt the best-fit results from $pexrav$.
Utilizing results from the other models however would not alter the conclusions of this work. 
Further note that in $relxill$, $relxillCp$ and $xillverCp$, the reflection component and the Fe lines are coupled under certain assumptions, which may bias the \ec (or $T_{\rm e}$) measurements in some sources \citep[e.g.,][]{Zhangjx2018, Kang_2020}.

\begin{figure*}
\includegraphics[width=2.42in]{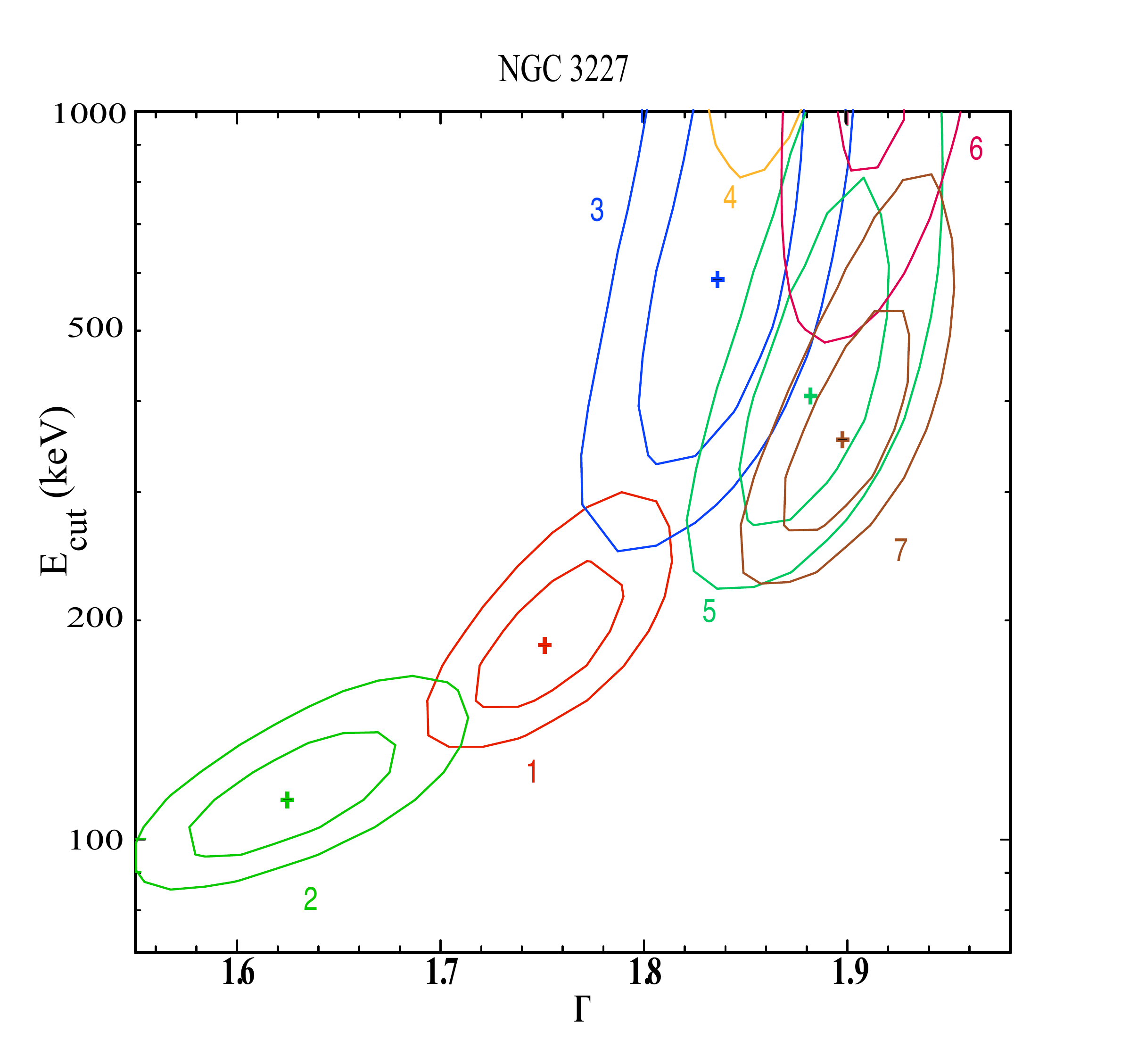}
\includegraphics[width=2.42in]{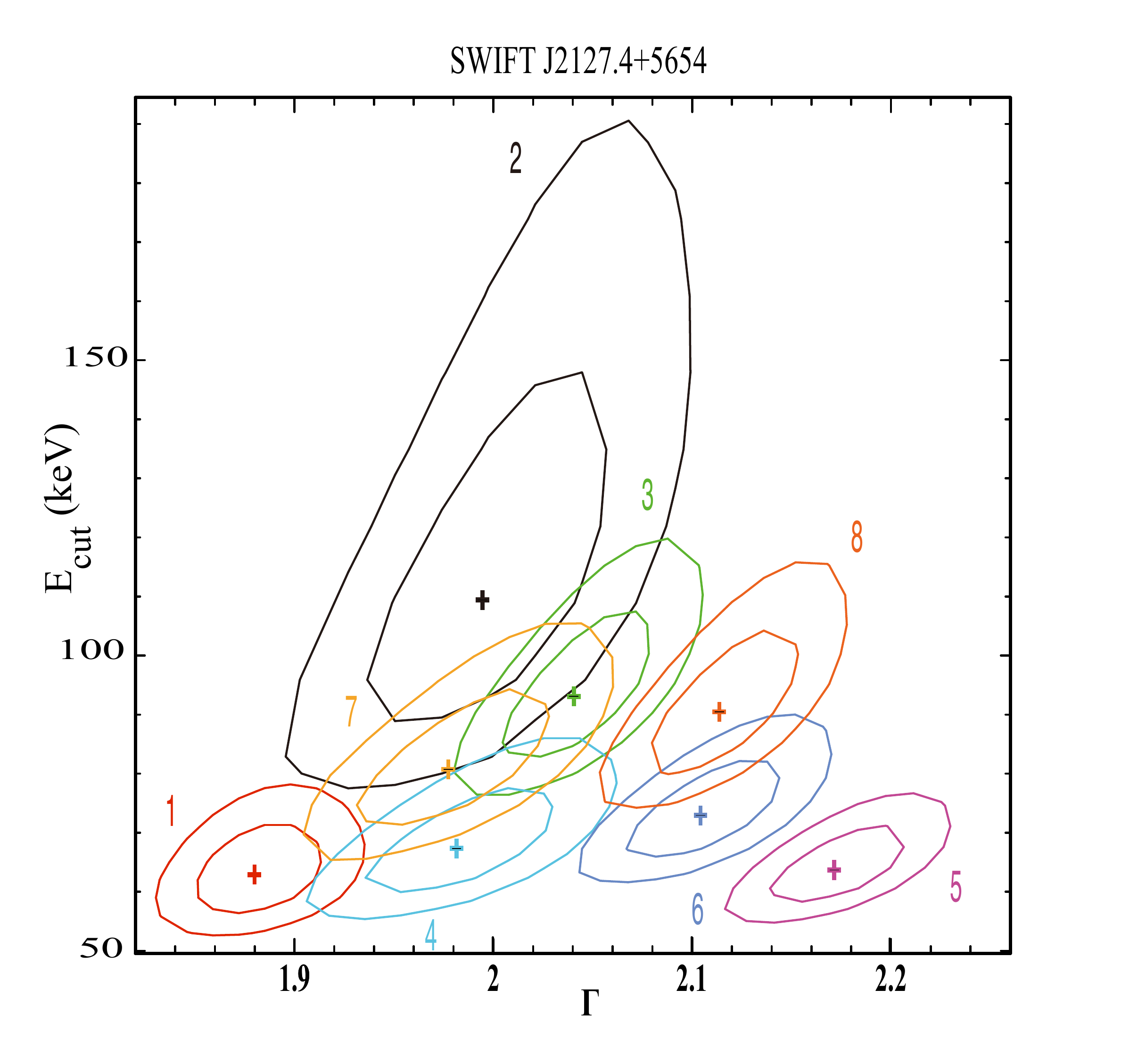}
\includegraphics[width=2.12 in]{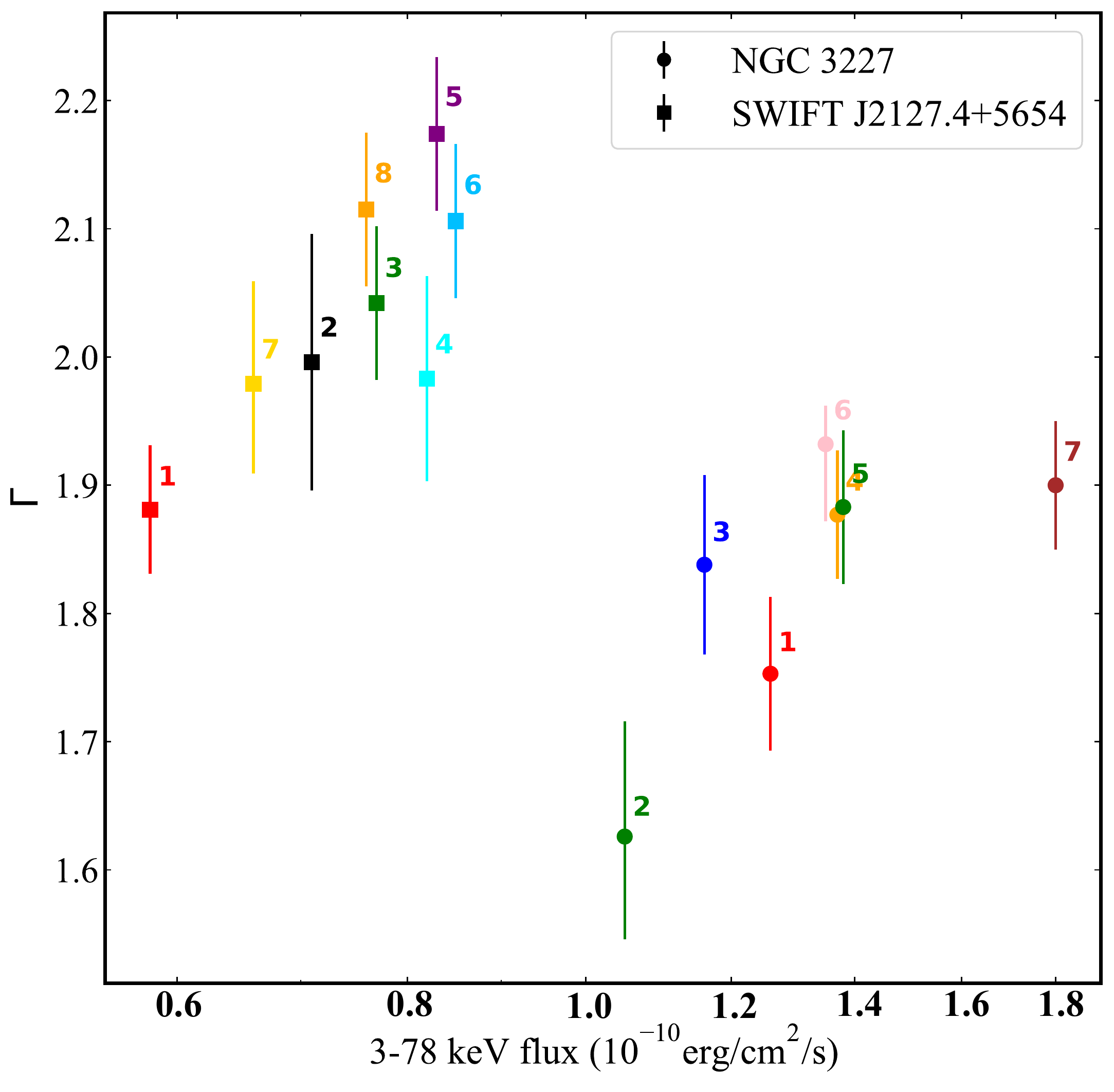}
\caption{Contour plots (at 1 $\sigma$ and 90\% confidence levels) of $\Gamma$ vs. $E_{\rm cut}$ ($\Gamma$ and $E_{\rm pexrav}$ from Tab. \ref{tab:results}), and the $\Gamma$ vs. 3 -- 78 keV flux variabilities. 
The observations are sorted and numbered by observation date (see Tab. \ref{tab:results}).
 }
\label{fig:contour}
\end{figure*}

We plot the $E_{\rm cut}$ vs. $\Gamma$ contours in Fig. \ref{fig:contour} to illustrate the \ec variation patterns in two sources. 
A clear positive correlation between $E_{\rm cut}$ and $\Gamma$ is seen in NGC 3227. However the $E_{\rm cut}$--$\Gamma$ plot of \swift exhibits a 
distinct $\Lambda$ shape: $E_{\rm cut}$ increases with $\Gamma$ at $\Gamma$ $\lesssim$ 2.05, but reversely decreases at $\Gamma$ $\gtrsim$ 2.05.
Though considering the degeneracy between $E_{\rm cut}$ and $\Gamma$, the rising part of the $\Lambda$ shape is less significant comparing with the declining part,
the overall variation trend in \swift clearly deviates from a monotonous function.
Meanwhile, the common ``softer-when-brighter'' trend in Seyfert galaxies is clear in both sources (Fig. \ref{fig:contour}).

\par To quantifying the significance of the $\Lambda$ shape, we perform Spearman rank-order correlation analyses on the rising (obs. No. 1, 2, 3, 4, 7) and descending (obs. No. 2, 3, 5, 6, 8) branches of the $\Lambda$ shape.
We find a positive correlation between $E_{\rm cut}$ and $\Gamma$ ($\rho = 0.87$ with a p-value = 0.054) for the rising branch, and a negative correlation for the descending branch (a $\rho = -0.90$ with a p-value = 0.037). We also perform linear regression to measure the slopes of the two branches of the $\Lambda$ shape ($\beta = 220$ for the left and $\beta = -240$ for the right).
However, the Spearman's correlation could be unreliable when the sample size is small (e.g., n < 10), and random fluctuation can produce strong correlation/anti-correlation in small samples. Meanwhile, the measurement errors and the degeneracy of the parameters should also be taken into account. 
We perform simulations to address these issues. 
Assuming there is no intrinsic \ec variation in SWIFT J2127.4+5654, we jointly fit all eight observations and derive a \ec of 80 keV. 
Starting from \ec = 80 keV and other best-fit spectral parameters (derived with \ec tied) for each observation, we create one artificial spectrum for each exposure using  $fakeit$.
We then perform spectral fitting to those faked spectra to measure the simulated \ec and $\Gamma$ and perform Spearman rank and linear regression analyses. 
We repeat the process 1000 times, and find only 26 runs out of them show stronger Spearman's correlation and steeper linear regression slope (compared with the observed values) for the left side of the $\Lambda$ shape, while only 2 runs out of 1000 for the right one.
This indicate the rising and descending of the $\Lambda$ pattern have statistical confidence level of 97.4\% and 99.8\% respectively, showing statistical fluctuations of the parameters and the degeneracy between $\Gamma$ and \ec are unlikely able to reproduce the observed $\Lambda$ pattern.

\section{Discussion} \label{sec:discussion}

\begin{figure}
\includegraphics[width=3.2in]{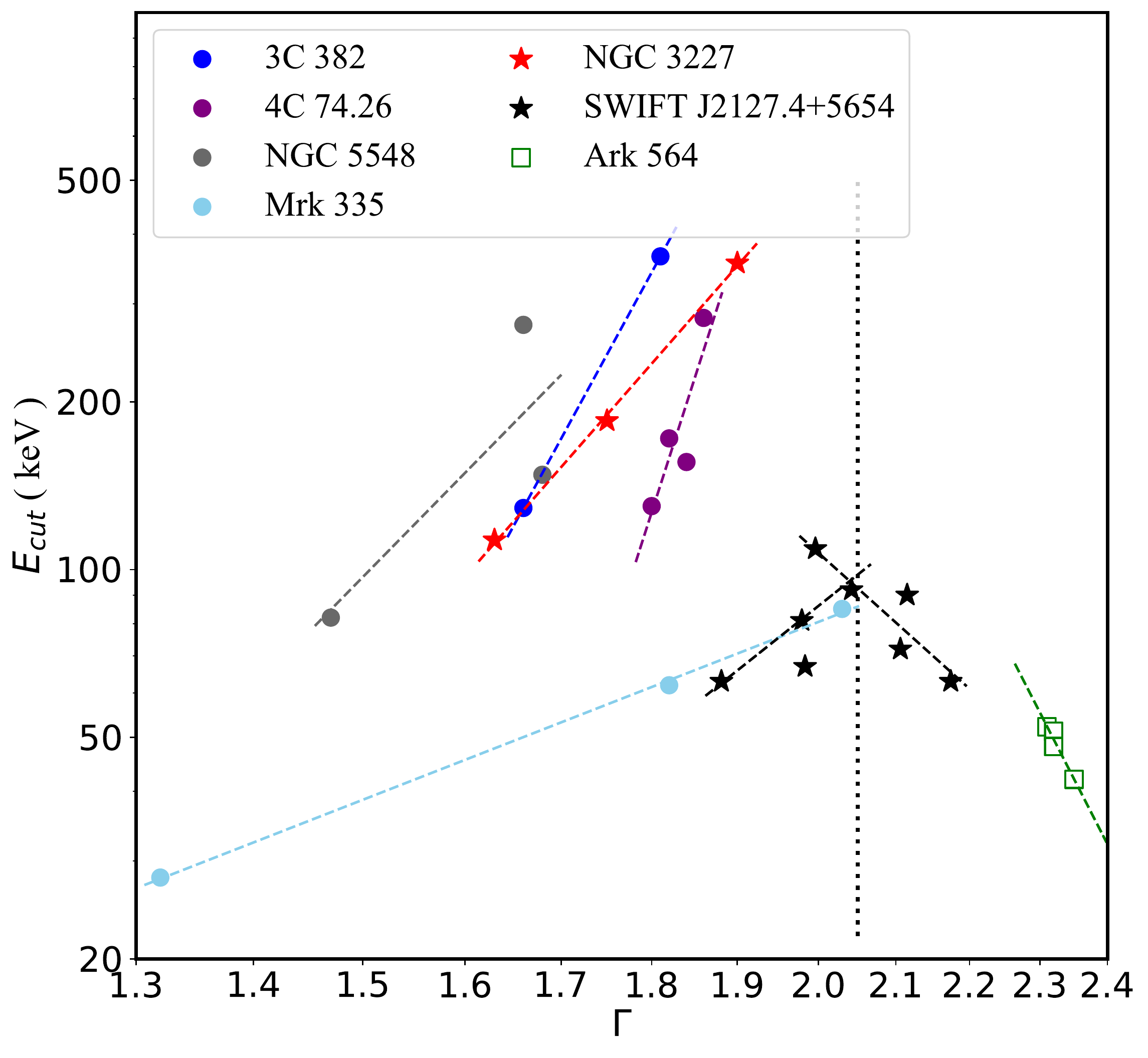}\\
\caption{The $\Gamma$ -- $E_{\rm cut}$ variation patterns of our two sources and those reported in literature. 
To avoid confusion, no error bars or contours are given, and we drop those exposures in which $E_{\rm cut}$ are non-detected. 
Besides, we dropped the data of 60002044006, NGC 5548, to avoid the potential photon index discrepancy issue, caused by XMM-Newton data as discussed in \S \ref{sec:obs}. 
A single best-fit direct line is over-plotted to demonstrate the overall variation trend of each source, except that for \swift we plot two lines to illustrate the $\Lambda$ shape. The vertical dotted line marks the transition to the left (right) of which \ec positively (negatively) correlates with $\Gamma$.
 }
\label{fig:all_gamma_E}
\end{figure}

In Fig. \ref{fig:all_gamma_E} we plot $E_{\rm cut}$ -- $\Gamma$ for both our sources together with those introduced in \S\ref{sec:intro}. For Ark 564 we convert the corona temperature $kT_{\rm e}$ given by \cite{Barua_2020} into $E_{\rm cut}$, assuming an optically thick corona and $kT_e \sim E_{\rm cut}/3$ \citep{Petrucci_2001}. For the other four sources we simply take measurements from \citet{Zhangjx2018}.
Though all of them follow the common ``softer-when-brighter'' trend  \citep[see Fig. \ref{fig:contour} and][]{Zhangjx2018, Barua_2020}, they show different $E_{\rm cut}$--$\Gamma$ variation patterns.
Interestingly, it appears that all seven sources could be unified with the $\Lambda$ shaped pattern seen in SWIFT J2127.4+5654, e.g., ``hotter-when-softer/brighter'' at $\Gamma$ $\lesssim$ 2.05, but ``cooler-when-softer/brighter'' at $\Gamma$ $\gtrsim$ 2.05, though \swift is the only source with $\Gamma$ varying across the break point, thus the only one showing the complete $\Lambda$ pattern in a single source. 

The common ``softer-when-brighter'' trend in Seyfert galaxies was generally attributed to presumbly cooler corona during brighter phases due to more effective cooling by more seed photons. However, such scenario clearly contradicts the discovery of ``hotter-when-brighter'' pattern detected in AGNs \citep[e.g.][]{Keek_2016, Zhangjx2018}. 
Geometry changes of the corona are required to reproduce the ``softer-when-brighter'' trend \citep[e.g.][]{Keek_2016,Zhangjx2018, Wu_2020}.
Specifically, the corona could be heated to a higher temperature and simultaneously driven to inflate during X-ray brighter phases \citep{Wu_2020}, leading to a smaller opacity and thus softer spectra, 
and reproducing the observed ``hotter-when-softer/brighter'' pattern (the rising part of the  $\Lambda$ pattern).
The inflation could be primarily vertical, as there are evidences that suggest the corona is vertically outflowing \citep[e.g.][]{liu2014} and the corona could reach higher heights during brighter phases \citep[][]{Wilkins_2015,Alston_2020}. Note in case of outflowing corona, if the outflowing velocity is higher during X-ray brighter phases, higher $E_{\rm cut}$ is also expected due to stronger Doppler shift.

\par As the X-ray flux brightening, spectrum softening and corona inflation continue, more seed photons from the disk could be intercepted, leading to higher cooling efficiency. Moreover, the cooling efficiency could be further boosted if the steeper X-ray spectrum is accompanied by a stronger soft X-ray excess component from the presumed warm corona \citep[e.g.][]{Petrucci2013} which could also contribute as seed photons. The latter mechanism may be essential, that the Compton cooling effect begins to dominate beyond a certain $\Gamma$, yielding the declining part of the  $\Lambda$ pattern (``cooler-when-brighter''). However, as \swift is yet the only individual source showing a $\Lambda$ pattern,  it is unclear whether the break point of $\Gamma$ $\sim$ 2.05 is universal, and if yes, why so. 

\par Meanwhile, pair production may also play a key role. As shown in \citet{Fabian_2015}, the coronae in many AGNs lie close to the electron--positron pair runaway line in the compactness--temperature diagram, suggesting the corona temperature is controlled by pair production. An inflated corona with a smaller compactness means a higher temperature limit by the run away pair-production. Moreover, \citet{Ghisellini_1994} showed that, in a pair-dominated corona with a certain compactness, the temperature positively correlates with $\Gamma$ for electron temperature $kT_e < m_e c^2$. 
Therefore positive correlations between X-ray flux and $E_{\rm cut}$ and between $\Gamma$ and $E_{\rm cut}$ are expected in pair-dominated corona. 

\par It is known that the NLS1 Ark 564 lies well below the thermal pair-production limit \citep{Kara_2017}, meanwhile 3C 382, 4C 74.26, NGC 5548 and Mrk 335 lie close to the limit \citep[e.g.][]{Zhangjx2018}.
Following \citet{Fabian_2015} we calculate the compactness, $l = 4\pi (m_p / m_e)(r_g / r)(L / L_{\rm edd})$, and dimensionless temperature, $\Theta = kT_e / m_e c^2$, for the two sources reported in this work,
where $kT_e \approx E_{\rm cut}/2$ {\footnote{{ We adopt  $kT_e \approx E_{\rm cut}/2$ to stay consistent with \citet{Fabian_2015} and \citet{Zhangjx2018}. Directly adopting the $kT_{\rm relxillCp}$ or $kT_{\rm xillverCp}$ we derived, or assuming $kT_e \approx E_{\rm cut}/3$ will not alter the main results presented here.}}. We adopt $r = 10$  $r_g$ and the 0.1--200 keV luminosity, $M=1.4 \times 10^7 M_{\odot}$ for NGC 3227 \citep{Graham_2008}, and $M=1.5 \times 10^7 M_{\odot}$ for SWIFT J2127.4+5654 \citep{Malizia_2008}. Comparing the results with the runaway pair production boundary in \citet{Stern_1995}, we find while \NGC lies close to or on the boundary, the NLS1 \swift also lie clearly below the boundary (though closer to the boundary compared with Ark 564, Fig. \ref{pairlimit}). 

\begin{figure}
\includegraphics[width=3.2in]{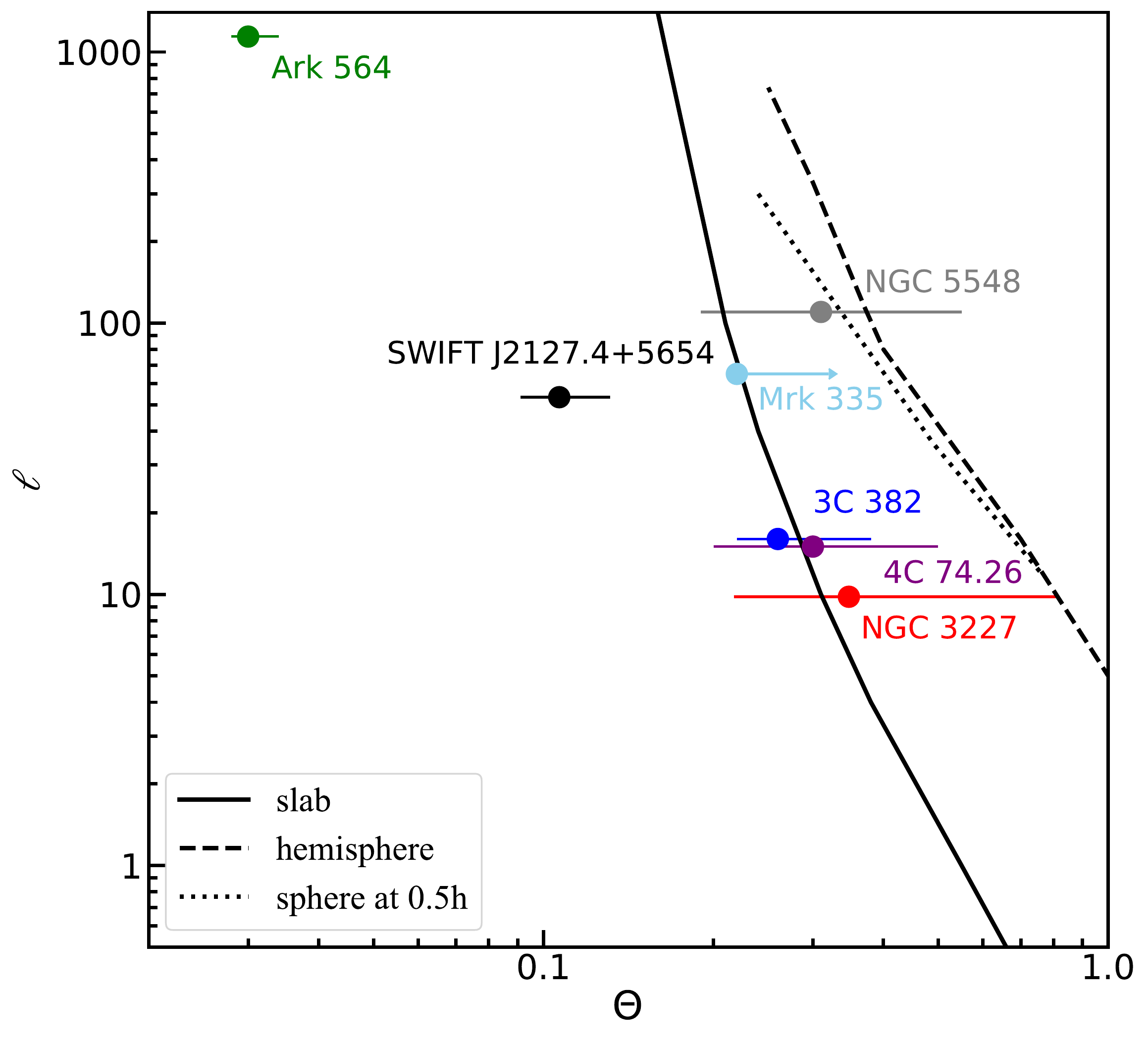}\\
\caption{ The $l$--$\Theta$ diagram of all seven sources. The results of Ark 564 in \citet{Kara_2017} as well as the four sources in \citet{Zhangjx2018} are simply taken and plotted. The maximum temperature that can be reached by a plasma dominated by the runaway pair production for three geometries \citep{Stern_1995} are over-plotted as lines. 
For the four sources from \citet{Zhangjx2018} we take the data points from its Fig. 13 (the data point with the higher $\Theta$ for each source). For \NGC and \swift we plot the observation with the highest detected $E_{\rm cut}$. For simplicity, no uncertainties to $l$ are given.
 }
\label{pairlimit}
\end{figure}

\par It is remarkable to note that there also likely exists a link between $E_{\rm cut}$ variation patterns and pair-dominance: pair-dominated coronae (those lying close to or on the pair limit in the $l$--$\Theta$ diagram) only exhibit ``hotter-when-softer/brighter'' trend, while a ``cooler-when-softer/brighter'' or $\Lambda$ shape is only possible in coronae which are not pair-dominated. In this case the pair production could counter against the higher cooling efficiency of an expanded and less compact corona in many AGNs.  Consequently the ``cooler-when-softer/brighter'' trend or the $\Lambda$ pattern only exists in sources lacking strong pair production, like the NLS1 SWIFT J2127.4+5654 and Ark 564. 
We note that \citet{Fabian2017} proposed that the coronae lying well below the pair limit (such as in Ark 564) could also be pair dominated if containing both thermal and non-thermal particles. 
However, if the possible link aforementioned does exist, it suggests the pair production in the cool coronae of Ark 564 and \swift is indeed weak.

\par Though the sample discussed above is small, the discoveries presented in this work shed new light on the coronal physics in AGNs. 
Future observations of $E_{\rm cut}$ variations in a larger sample of AGNs are desired to testify the universality of the  $\Lambda$ pattern, confirm the potential link between $E_{\rm cut}$ variation pattern and pair production, and independently probe the coronal physics.

\section*{Acknowledgements}

  This research has made use of the NuSTAR Data Analysis Software (NuSTARDAS) jointly developed by the ASI Science Data Center (ASDC, Italy) and the California Institute of Technology (USA). The work is supported by National Natural Science Foundation of China (grants No. 11421303, 11890693 $\&$ 12033006) and CAS Frontier Science Key Research Program (QYZDJ-SSW-SLH006).

\section*{Data Availability}
The data underlying this article are available in the article and in its online supplementary material.

\bibliographystyle{mnras}
\bibliography{mnras}

\begin{thebibliography}{}
\makeatletter
\relax
\def\mn@urlcharsother{\let\do\@makeother \do\$\do\&\do\#\do\^\do\_\do\%\do\~}
\def\mn@doi{\begingroup\mn@urlcharsother \@ifnextchar [ {\mn@doi@}
  {\mn@doi@[]}}
\def\mn@doi@[#1]#2{\def\@tempa{#1}\ifx\@tempa\@empty \href
  {http://dx.doi.org/#2} {doi:#2}\else \href {http://dx.doi.org/#2} {#1}\fi
  \endgroup}
\def\mn@eprint#1#2{\mn@eprint@#1:#2::\@nil}
\def\mn@eprint@arXiv#1{\href {http://arxiv.org/abs/#1} {{\tt arXiv:#1}}}
\def\mn@eprint@dblp#1{\href {http://dblp.uni-trier.de/rec/bibtex/#1.xml}
  {dblp:#1}}
\def\mn@eprint@#1:#2:#3:#4\@nil{\def\@tempa {#1}\def\@tempb {#2}\def\@tempc
  {#3}\ifx \@tempc \@empty \let \@tempc \@tempb \let \@tempb \@tempa \fi \ifx
  \@tempb \@empty \def\@tempb {arXiv}\fi \@ifundefined
  {mn@eprint@\@tempb}{\@tempb:\@tempc}{\expandafter \expandafter \csname
  mn@eprint@\@tempb\endcsname \expandafter{\@tempc}}}

\bibitem[\protect\citeauthoryear{{Alston} et~al.,}{{Alston}
  et~al.}{2020}]{Alston_2020}
{Alston} W.~N.,  et~al., 2020, \mn@doi [Nature Astronomy]
  {10.1038/s41550-019-1002-x}, \href
  {https://ui.adsabs.harvard.edu/abs/2020NatAs.tmp....2A} {p.~2}

\bibitem[\protect\citeauthoryear{Anders \& Grevesse}{Anders \&
  Grevesse}{1989}]{ANDERS1989197}
Anders E.,  Grevesse N.,  1989, \mn@doi [Geochimica et Cosmochimica Acta]
  {10.1016/0016-7037(89)90286-X}, 53, 197

\bibitem[\protect\citeauthoryear{{Arnaud}}{{Arnaud}}{1996}]{Arnaud_1996}
{Arnaud} K.~A.,  1996, in {Jacoby} G.~H.,  {Barnes} J.,  eds,  Astronomical
  Society of the Pacific Conference Series Vol. 101, Astronomical Data Analysis
  Software and Systems V. p.~17

\bibitem[\protect\citeauthoryear{Ballantyne et~al.,}{Ballantyne
  et~al.}{2014}]{Ballantyne_2014}
Ballantyne D.~R.,  et~al., 2014, \mn@doi [The Astrophysical Journal]
  {10.1088/0004-637x/794/1/62}, 794, 62

\bibitem[\protect\citeauthoryear{{Barua}, {Jithesh}, {Misra}, {Dewangan},
  {Sarma}  \& {Pathak}}{{Barua} et~al.}{2020}]{Barua_2020}
{Barua} S.,  {Jithesh} V.,  {Misra} R.,  {Dewangan} G.~C.,  {Sarma} R.,
  {Pathak} A.,  2020, \mn@doi [\mnras] {10.1093/mnras/staa067}, \href
  {https://ui.adsabs.harvard.edu/abs/2020MNRAS.492.3041B} {492, 3041}

\bibitem[\protect\citeauthoryear{{Cappi} et~al.,}{{Cappi}
  et~al.}{2016}]{gamma_xmm_1}
{Cappi} M.,  et~al., 2016, \mn@doi [\aap] {10.1051/0004-6361/201628464}, \href
  {https://ui.adsabs.harvard.edu/abs/2016A&A...592A..27C} {592, A27}

\bibitem[\protect\citeauthoryear{{Dauser}, {Wilms}, {Reynolds}  \&
  {Brenneman}}{{Dauser} et~al.}{2010}]{Dauser_2010}
{Dauser} T.,  {Wilms} J.,  {Reynolds} C.~S.,   {Brenneman} L.~W.,  2010,
  \mn@doi [\mnras] {10.1111/j.1365-2966.2010.17393.x}, \href
  {https://ui.adsabs.harvard.edu/abs/2010MNRAS.409.1534D} {409, 1534}

\bibitem[\protect\citeauthoryear{{Fabian}, {Lohfink}, {Kara}, {Parker},
  {Vasudevan}  \& {Reynolds}}{{Fabian} et~al.}{2015}]{Fabian_2015}
{Fabian} A.~C.,  {Lohfink} A.,  {Kara} E.,  {Parker} M.~L.,  {Vasudevan} R.,
  {Reynolds} C.~S.,  2015, \mn@doi [\mnras] {10.1093/mnras/stv1218}, \href
  {https://ui.adsabs.harvard.edu/abs/2015MNRAS.451.4375F} {451, 4375}

\bibitem[\protect\citeauthoryear{{Fabian}, {Lohfink}, {Belmont}, {Malzac}  \&
  {Coppi}}{{Fabian} et~al.}{2017}]{Fabian2017}
{Fabian} A.~C.,  {Lohfink} A.,  {Belmont} R.,  {Malzac} J.,   {Coppi} P.,
  2017, \mn@doi [\mnras] {10.1093/mnras/stx221}, \href
  {https://ui.adsabs.harvard.edu/abs/2017MNRAS.467.2566F} {467, 2566}

\bibitem[\protect\citeauthoryear{{Garc{\'\i}a} et~al.,}{{Garc{\'\i}a}
  et~al.}{2014}]{Garc_2014}
{Garc{\'\i}a} J.,  et~al., 2014, \mn@doi [\apj] {10.1088/0004-637X/782/2/76},
  \href {https://ui.adsabs.harvard.edu/abs/2014ApJ...782...76G} {782, 76}

\bibitem[\protect\citeauthoryear{{Ghisellini} \& {Haardt}}{{Ghisellini} \&
  {Haardt}}{1994}]{Ghisellini_1994}
{Ghisellini} G.,  {Haardt} F.,  1994, \mn@doi [\apjl] {10.1086/187411}, \href
  {https://ui.adsabs.harvard.edu/abs/1994ApJ...429L..53G} {429, L53}

\bibitem[\protect\citeauthoryear{{Graham}}{{Graham}}{2008}]{Graham_2008}
{Graham} A.~W.,  2008, \mn@doi [\pasa] {10.1071/AS08013}, \href
  {https://ui.adsabs.harvard.edu/abs/2008PASA...25..167G} {25, 167}

\bibitem[\protect\citeauthoryear{{Haardt} \& {Maraschi}}{{Haardt} \&
  {Maraschi}}{1991}]{Haardt_1991}
{Haardt} F.,  {Maraschi} L.,  1991, \mn@doi [\apjl] {10.1086/186171}, \href
  {https://ui.adsabs.harvard.edu/abs/1991ApJ...380L..51H} {380, L51}

\bibitem[\protect\citeauthoryear{{Haardt}, {Maraschi}  \&
  {Ghisellini}}{{Haardt} et~al.}{1994}]{Haardt_1994}
{Haardt} F.,  {Maraschi} L.,   {Ghisellini} G.,  1994, \mn@doi [\apjl]
  {10.1086/187520}, \href
  {https://ui.adsabs.harvard.edu/abs/1994ApJ...432L..95H} {432, L95}

\bibitem[\protect\citeauthoryear{Harrison et~al.,}{Harrison
  et~al.}{2013}]{Harrison_2013}
Harrison F.~A.,  et~al., 2013, \mn@doi [The Astrophysical Journal]
  {10.1088/0004-637x/770/2/103}, 770, 103

\bibitem[\protect\citeauthoryear{Kang, Wang  \& Kang}{Kang
  et~al.}{2020}]{Kang_2020}
Kang J.,  Wang J.,   Kang W.,  2020, \mn@doi [The Astrophysical Journal]
  {10.3847/1538-4357/abadf5}, 901, 111

\bibitem[\protect\citeauthoryear{{Kara}, {Garc{\'\i}a}, {Lohfink}, {Fabian},
  {Reynolds}, {Tombesi}  \& {Wilkins}}{{Kara} et~al.}{2017}]{Kara_2017}
{Kara} E.,  {Garc{\'\i}a} J.~A.,  {Lohfink} A.,  {Fabian} A.~C.,  {Reynolds}
  C.~S.,  {Tombesi} F.,   {Wilkins} D.~R.,  2017, \mn@doi [\mnras]
  {10.1093/mnras/stx792}, \href
  {https://ui.adsabs.harvard.edu/abs/2017MNRAS.468.3489K} {468, 3489}

\bibitem[\protect\citeauthoryear{{Keek} \& {Ballantyne}}{{Keek} \&
  {Ballantyne}}{2016}]{Keek_2016}
{Keek} L.,  {Ballantyne} D.~R.,  2016, \mn@doi [\mnras]
  {10.1093/mnras/stv2882}, \href
  {https://ui.adsabs.harvard.edu/abs/2016MNRAS.456.2722K} {456, 2722}

\bibitem[\protect\citeauthoryear{{Liu}, {Wang}, {Yang}, {Zhu}  \& {Zhou}}{{Liu}
  et~al.}{2014}]{liu2014}
{Liu} T.,  {Wang} J.-X.,  {Yang} H.,  {Zhu} F.-F.,   {Zhou} Y.-Y.,  2014,
  \mn@doi [\apj] {10.1088/0004-637X/783/2/106}, \href
  {https://ui.adsabs.harvard.edu/abs/2014ApJ...783..106L} {783, 106}

\bibitem[\protect\citeauthoryear{{Madsen} et~al.,}{{Madsen}
  et~al.}{2015}]{Madsen_2015}
{Madsen} K.~K.,  et~al., 2015, \mn@doi [\apjs] {10.1088/0067-0049/220/1/8},
  \href {https://ui.adsabs.harvard.edu/abs/2015ApJS..220....8M} {220, 8}

\bibitem[\protect\citeauthoryear{{Madsen}, {Grefenstette}, {Pike}, {Miyasaka},
  {Brightman}, {Forster}  \& {Harrison}}{{Madsen} et~al.}{2020}]{Madsen_2020}
{Madsen} K.~K.,  {Grefenstette} B.~W.,  {Pike} S.,  {Miyasaka} H.,  {Brightman}
  M.,  {Forster} K.,   {Harrison} F.~A.,  2020, arXiv e-prints, \href
  {https://ui.adsabs.harvard.edu/abs/2020arXiv200500569M} {p. arXiv:2005.00569}

\bibitem[\protect\citeauthoryear{{Magdziarz} \& {Zdziarski}}{{Magdziarz} \&
  {Zdziarski}}{1995}]{Magdziarz_1995}
{Magdziarz} P.,  {Zdziarski} A.~A.,  1995, \mn@doi [\mnras]
  {10.1093/mnras/273.3.837}, \href
  {https://ui.adsabs.harvard.edu/abs/1995MNRAS.273..837M} {273, 837}

\bibitem[\protect\citeauthoryear{{Malizia} et~al.,}{{Malizia}
  et~al.}{2008}]{Malizia_2008}
{Malizia} A.,  et~al., 2008, \mn@doi [\mnras]
  {10.1111/j.1365-2966.2008.13657.x}, \href
  {https://ui.adsabs.harvard.edu/abs/2008MNRAS.389.1360M} {389, 1360}

\bibitem[\protect\citeauthoryear{{Marinucci} et~al.,}{{Marinucci}
  et~al.}{2014}]{Marinucci_2014}
{Marinucci} A.,  et~al., 2014, \mn@doi [\mnras] {10.1093/mnras/stu404}, \href
  {https://ui.adsabs.harvard.edu/abs/2014MNRAS.440.2347M} {440, 2347}

\bibitem[\protect\citeauthoryear{{Markowitz}, {Edelson}  \&
  {Vaughan}}{{Markowitz} et~al.}{2003}]{Markowitz_2003}
{Markowitz} A.,  {Edelson} R.,   {Vaughan} S.,  2003, \mn@doi [\apj]
  {10.1086/379103}, \href
  {https://ui.adsabs.harvard.edu/abs/2003ApJ...598..935M} {598, 935}

\bibitem[\protect\citeauthoryear{{Markowitz}, {Reeves}, {George}, {Braito},
  {Smith}, {Vaughan}, {Ar{\'e}valo}  \& {Tombesi}}{{Markowitz}
  et~al.}{2009}]{Markowitz_2009}
{Markowitz} A.,  {Reeves} J.~N.,  {George} I.~M.,  {Braito} V.,  {Smith} R.,
  {Vaughan} S.,  {Ar{\'e}valo} P.,   {Tombesi} F.,  2009, \mn@doi [\apj]
  {10.1088/0004-637X/691/2/922}, \href
  {https://ui.adsabs.harvard.edu/abs/2009ApJ...691..922M} {691, 922}

\bibitem[\protect\citeauthoryear{{Matt} et~al.,}{{Matt}
  et~al.}{2015}]{Matt_2015_highE}
{Matt} G.,  et~al., 2015, \mn@doi [\mnras] {10.1093/mnras/stu2653}, \href
  {https://ui.adsabs.harvard.edu/abs/2015MNRAS.447.3029M} {447, 3029}

\bibitem[\protect\citeauthoryear{{Middei} et~al.,}{{Middei}
  et~al.}{2019a}]{Middei_2019}
{Middei} R.,  et~al., 2019a, \mn@doi [\mnras] {10.1093/mnras/sty3379}, \href
  {https://ui.adsabs.harvard.edu/abs/2019MNRAS.483.4695M} {483, 4695}

\bibitem[\protect\citeauthoryear{{Middei} et~al.,}{{Middei}
  et~al.}{2019b}]{gamma_xmm_2}
{Middei} R.,  et~al., 2019b, \mn@doi [\mnras] {10.1093/mnras/sty3379}, \href
  {https://ui.adsabs.harvard.edu/abs/2019MNRAS.483.4695M} {483, 4695}

\bibitem[\protect\citeauthoryear{{Molina}, {Bassani}, {Malizia}, {Stephen},
  {Bird}, {Bazzano}  \& {Ubertini}}{{Molina} et~al.}{2013}]{Molina_2013}
{Molina} M.,  {Bassani} L.,  {Malizia} A.,  {Stephen} J.~B.,  {Bird} A.~J.,
  {Bazzano} A.,   {Ubertini} P.,  2013, \mn@doi [\mnras]
  {10.1093/mnras/stt844}, \href
  {https://ui.adsabs.harvard.edu/abs/2013MNRAS.433.1687M} {433, 1687}

\bibitem[\protect\citeauthoryear{Molina, Malizia, Bassani, Ursini, Bazzano  \&
  Ubertini}{Molina et~al.}{2019}]{Molina_2019}
Molina M.,  Malizia A.,  Bassani L.,  Ursini F.,  Bazzano A.,   Ubertini P.,
  2019, \mn@doi [Monthly Notices of the Royal Astronomical Society]
  {10.1093/mnras/stz156}, 484, 2735

\bibitem[\protect\citeauthoryear{{Panagiotou} \& {Walter}}{{Panagiotou} \&
  {Walter}}{2020}]{PW2020}
{Panagiotou} C.,  {Walter} R.,  2020, \mn@doi [\aap]
  {10.1051/0004-6361/201937390}, \href
  {https://ui.adsabs.harvard.edu/abs/2020A&A...640A..31P} {640, A31}

\bibitem[\protect\citeauthoryear{{Petrucci} et~al.,}{{Petrucci}
  et~al.}{2001}]{Petrucci_2001}
{Petrucci} P.~O.,  et~al., 2001, \mn@doi [\apj] {10.1086/321629}, \href
  {https://ui.adsabs.harvard.edu/abs/2001ApJ...556..716P} {556, 716}

\bibitem[\protect\citeauthoryear{{Petrucci} et~al.,}{{Petrucci}
  et~al.}{2013}]{Petrucci2013}
{Petrucci} P.~O.,  et~al., 2013, \mn@doi [\aap] {10.1051/0004-6361/201219956},
  \href {https://ui.adsabs.harvard.edu/abs/2013A&A...549A..73P} {549, A73}

\bibitem[\protect\citeauthoryear{{Ponti} et~al.,}{{Ponti}
  et~al.}{2018}]{gamma_xmm_3}
{Ponti} G.,  et~al., 2018, \mn@doi [\mnras] {10.1093/mnras/stx2425}, \href
  {https://ui.adsabs.harvard.edu/abs/2018MNRAS.473.2304P} {473, 2304}

\bibitem[\protect\citeauthoryear{{Sanfrutos}, {Miniutti},
  {Ag{\'\i}s-Gonz{\'a}lez}, {Fabian}, {Miller}, {Panessa}  \&
  {Zoghbi}}{{Sanfrutos} et~al.}{2013}]{Sanfrutos_2013}
{Sanfrutos} M.,  {Miniutti} G.,  {Ag{\'\i}s-Gonz{\'a}lez} B.,  {Fabian} A.~C.,
  {Miller} J.~M.,  {Panessa} F.,   {Zoghbi} A.,  2013, \mn@doi [\mnras]
  {10.1093/mnras/stt1675}, \href
  {https://ui.adsabs.harvard.edu/abs/2013MNRAS.436.1588S} {436, 1588}

\bibitem[\protect\citeauthoryear{{Sarma}, {Tripathi}, {Misra}, {Dewangan},
  {Pathak}  \& {Sarma}}{{Sarma} et~al.}{2015}]{Sarma2015}
{Sarma} R.,  {Tripathi} S.,  {Misra} R.,  {Dewangan} G.,  {Pathak} A.,
  {Sarma} J.~K.,  2015, \mn@doi [\mnras] {10.1093/mnras/stv005}, \href
  {https://ui.adsabs.harvard.edu/abs/2015MNRAS.448.1541S} {448, 1541}

\bibitem[\protect\citeauthoryear{Shu, Yaqoob  \& Wang}{Shu
  et~al.}{2010}]{Shu_2010}
Shu X.~W.,  Yaqoob T.,   Wang J.~X.,  2010, \mn@doi [The Astrophysical Journal
  Supplement Series] {10.1088/0067-0049/187/2/581}, 187, 581

\bibitem[\protect\citeauthoryear{{Sobolewska} \& {Papadakis}}{{Sobolewska} \&
  {Papadakis}}{2009}]{Sobolewska_2009}
{Sobolewska} M.~A.,  {Papadakis} I.~E.,  2009, \mn@doi [\mnras]
  {10.1111/j.1365-2966.2009.15382.x}, \href
  {https://ui.adsabs.harvard.edu/abs/2009MNRAS.399.1597S} {399, 1597}

\bibitem[\protect\citeauthoryear{{Stern}, {Poutanen}, {Svensson}, {Sikora}  \&
  {Begelman}}{{Stern} et~al.}{1995}]{Stern_1995}
{Stern} B.~E.,  {Poutanen} J.,  {Svensson} R.,  {Sikora} M.,   {Begelman}
  M.~C.,  1995, \mn@doi [\apjl] {10.1086/309617}, \href
  {https://ui.adsabs.harvard.edu/abs/1995ApJ...449L..13S} {449, L13}

\bibitem[\protect\citeauthoryear{{Tortosa}, {Bianchi}, {Marinucci}, {Matt}  \&
  {Petrucci}}{{Tortosa} et~al.}{2018}]{Tortosa_2018}
{Tortosa} A.,  {Bianchi} S.,  {Marinucci} A.,  {Matt} G.,   {Petrucci} P.~O.,
  2018, \mn@doi [\aap] {10.1051/0004-6361/201732382}, \href
  {https://ui.adsabs.harvard.edu/abs/2018A&A...614A..37T} {614, A37}

\bibitem[\protect\citeauthoryear{{Turner}, {Reeves}, {Braito}, {Lobban},
  {Kraemer}  \& {Miller}}{{Turner} et~al.}{2018}]{Turner_2018}
{Turner} T.~J.,  {Reeves} J.~N.,  {Braito} V.,  {Lobban} A.,  {Kraemer} S.,
  {Miller} L.,  2018, \mn@doi [\mnras] {10.1093/mnras/sty2447}, \href
  {https://ui.adsabs.harvard.edu/abs/2018MNRAS.481.2470T} {481, 2470}

\bibitem[\protect\citeauthoryear{{Ursini} et~al.,}{{Ursini}
  et~al.}{2015}]{Ursini_2015}
{Ursini} F.,  et~al., 2015, \mn@doi [\aap] {10.1051/0004-6361/201425401}, \href
  {https://ui.adsabs.harvard.edu/abs/2015A&A...577A..38U} {577, A38}

\bibitem[\protect\citeauthoryear{{Ursini} et~al.,}{{Ursini}
  et~al.}{2016}]{Ursini_2016}
{Ursini} F.,  et~al., 2016, \mn@doi [\mnras] {10.1093/mnras/stw2022}, \href
  {https://ui.adsabs.harvard.edu/abs/2016MNRAS.463..382U} {463, 382}

\bibitem[\protect\citeauthoryear{{V{\'e}ron-Cetty} \&
  {V{\'e}ron}}{{V{\'e}ron-Cetty} \& {V{\'e}ron}}{2006}]{Classification}
{V{\'e}ron-Cetty} M.~P.,  {V{\'e}ron} P.,  2006, \mn@doi [\aap]
  {10.1051/0004-6361:20065177}, \href
  {https://ui.adsabs.harvard.edu/abs/2006A&A...455..773V} {455, 773}

\bibitem[\protect\citeauthoryear{Wik et~al.,}{Wik et~al.}{2014}]{Wik_2014}
Wik D.~R.,  et~al., 2014, \mn@doi [The Astrophysical Journal]
  {10.1088/0004-637x/792/1/48}, 792, 48

\bibitem[\protect\citeauthoryear{{Wilkins} \& {Gallo}}{{Wilkins} \&
  {Gallo}}{2015}]{Wilkins_2015}
{Wilkins} D.~R.,  {Gallo} L.~C.,  2015, \mn@doi [\mnras]
  {10.1093/mnras/stv162}, \href
  {https://ui.adsabs.harvard.edu/abs/2015MNRAS.449..129W} {449, 129}

\bibitem[\protect\citeauthoryear{{Wu}, {Wang}, {Cai}, {Kang}, {Liu}  \&
  {Cai}}{{Wu} et~al.}{2020}]{Wu_2020}
{Wu} Y.-J.,  {Wang} J.-X.,  {Cai} Z.-Y.,  {Kang} J.-L.,  {Liu} T.,   {Cai} Z.,
  2020, arXiv e-prints, \href
  {https://ui.adsabs.harvard.edu/abs/2020arXiv200803284W} {p. arXiv:2008.03284}

\bibitem[\protect\citeauthoryear{{Zdziarski}, {Poutanen}  \&
  {Johnson}}{{Zdziarski} et~al.}{2000}]{Zdziarski_2000}
{Zdziarski} A.~A.,  {Poutanen} J.,   {Johnson} W.~N.,  2000, \mn@doi [\apj]
  {10.1086/317046}, \href
  {https://ui.adsabs.harvard.edu/abs/2000ApJ...542..703Z} {542, 703}

\bibitem[\protect\citeauthoryear{{Zhang}, {Wang}  \& {Zhu}}{{Zhang}
  et~al.}{2018}]{Zhangjx2018}
{Zhang} J.-X.,  {Wang} J.-X.,   {Zhu} F.-F.,  2018, \mn@doi [\apj]
  {10.3847/1538-4357/aacf92}, \href
  {https://ui.adsabs.harvard.edu/abs/2018ApJ...863...71Z} {863, 71}

\bibitem[\protect\citeauthoryear{{Zoghbi} et~al.,}{{Zoghbi}
  et~al.}{2017}]{Zoghbi_2017}
{Zoghbi} A.,  et~al., 2017, \mn@doi [\apj] {10.3847/1538-4357/aa582c}, \href
  {https://ui.adsabs.harvard.edu/abs/2017ApJ...836....2Z} {836, 2}

\makeatother
\end{thebibliography}

\appendix

\section{NuSTAR spectra}

\begin{figure*}
\includegraphics[width=2.2in]{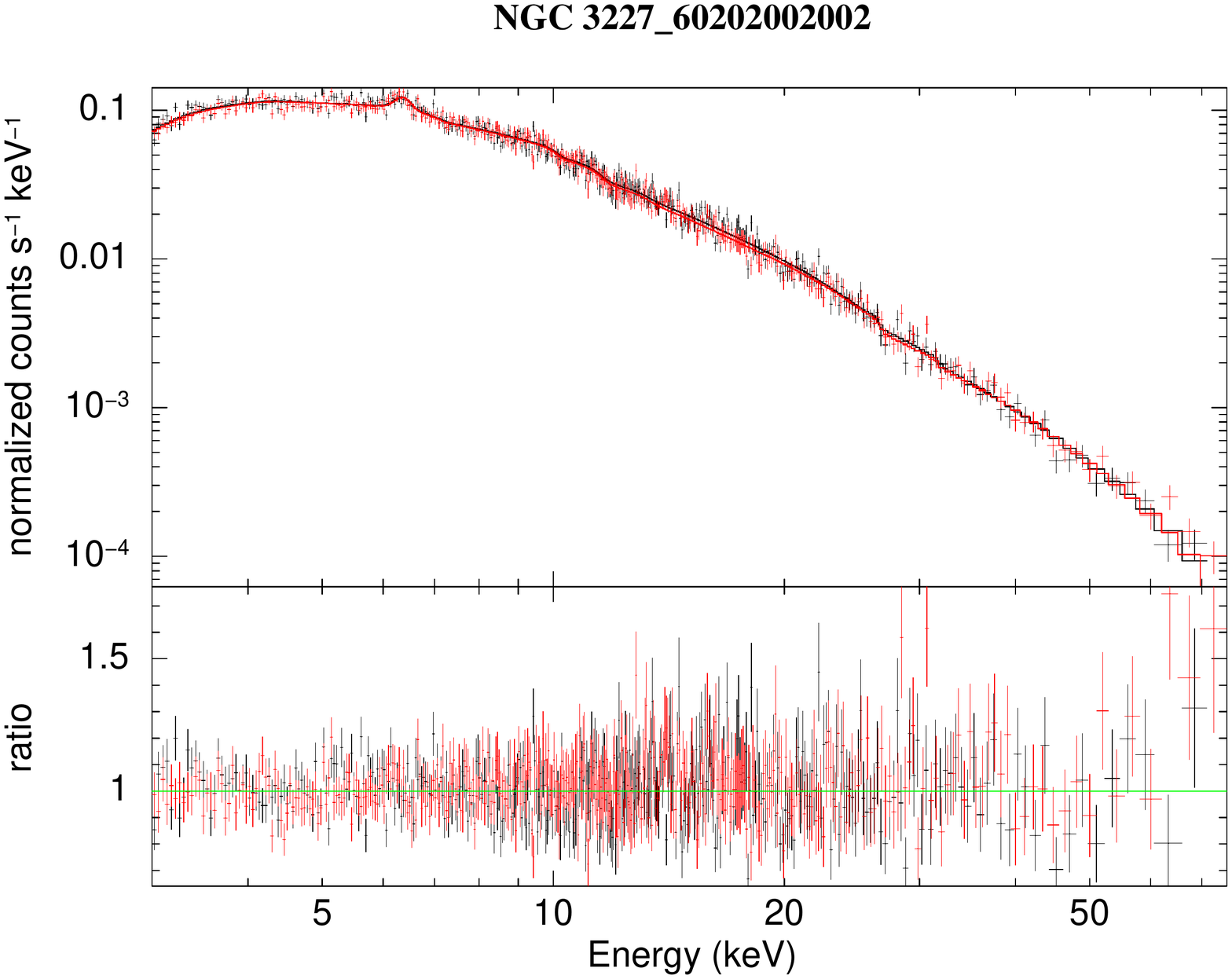}
\includegraphics[width=2.2in]{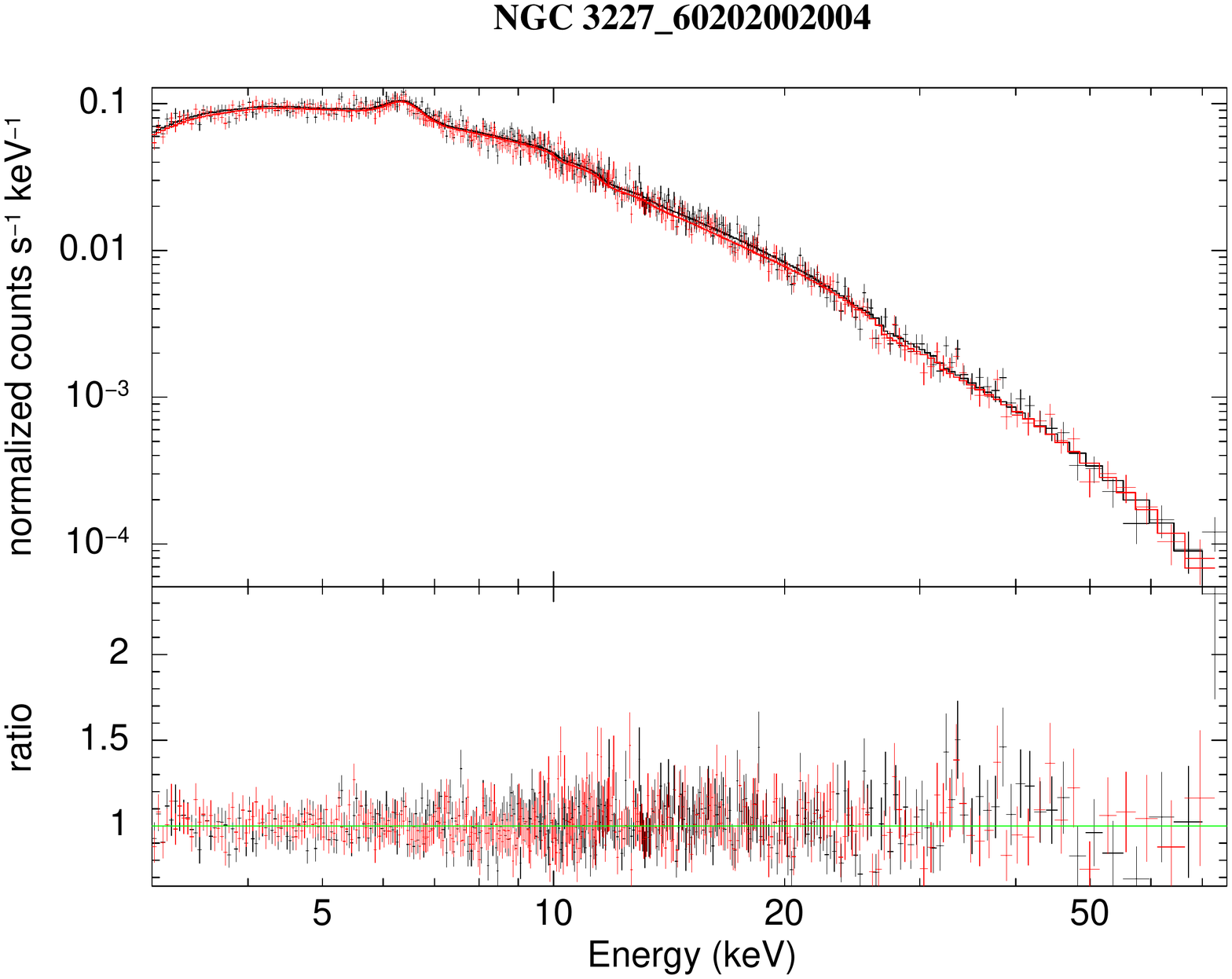}
\includegraphics[width=2.2in]{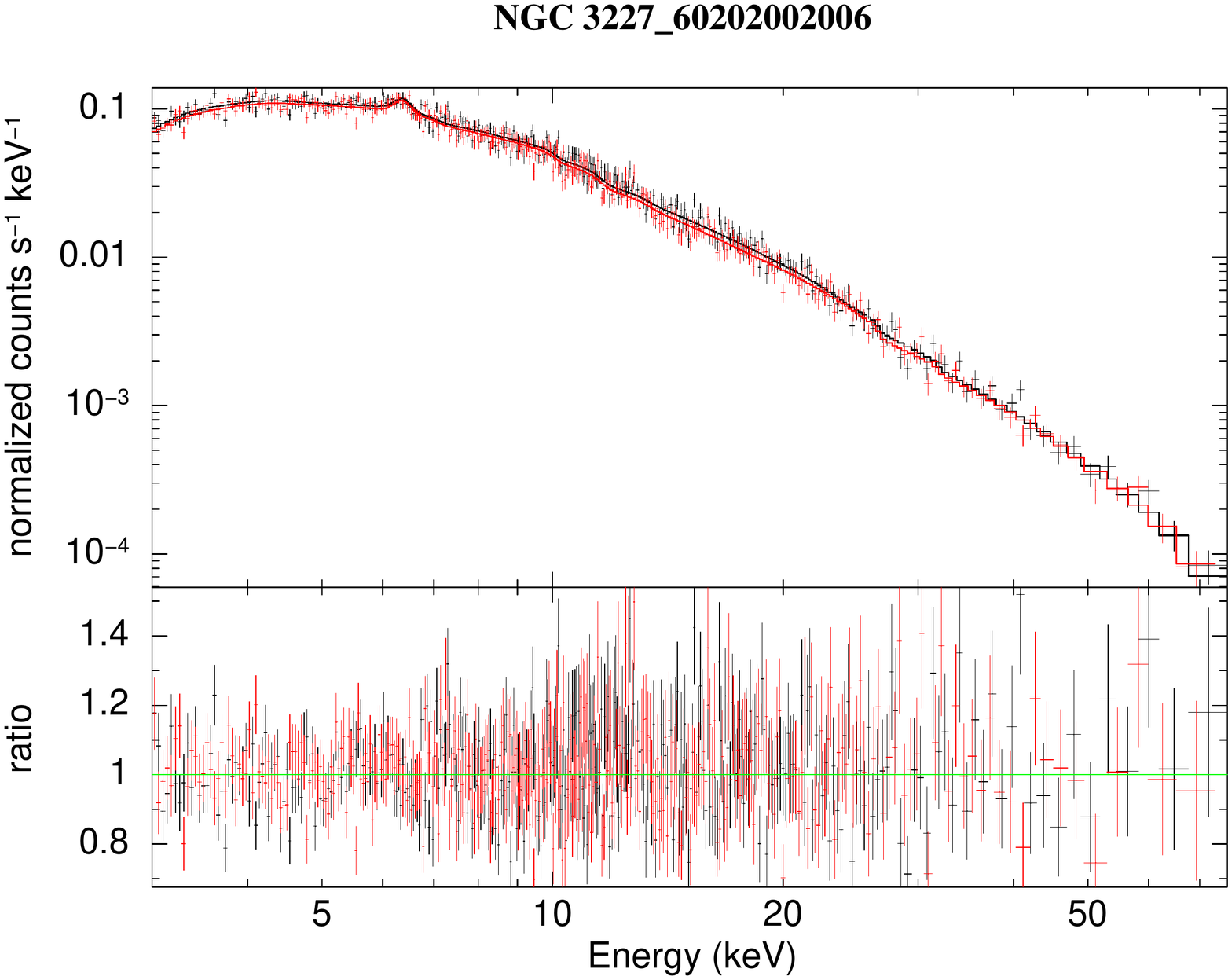}\\
\includegraphics[width=2.2in]{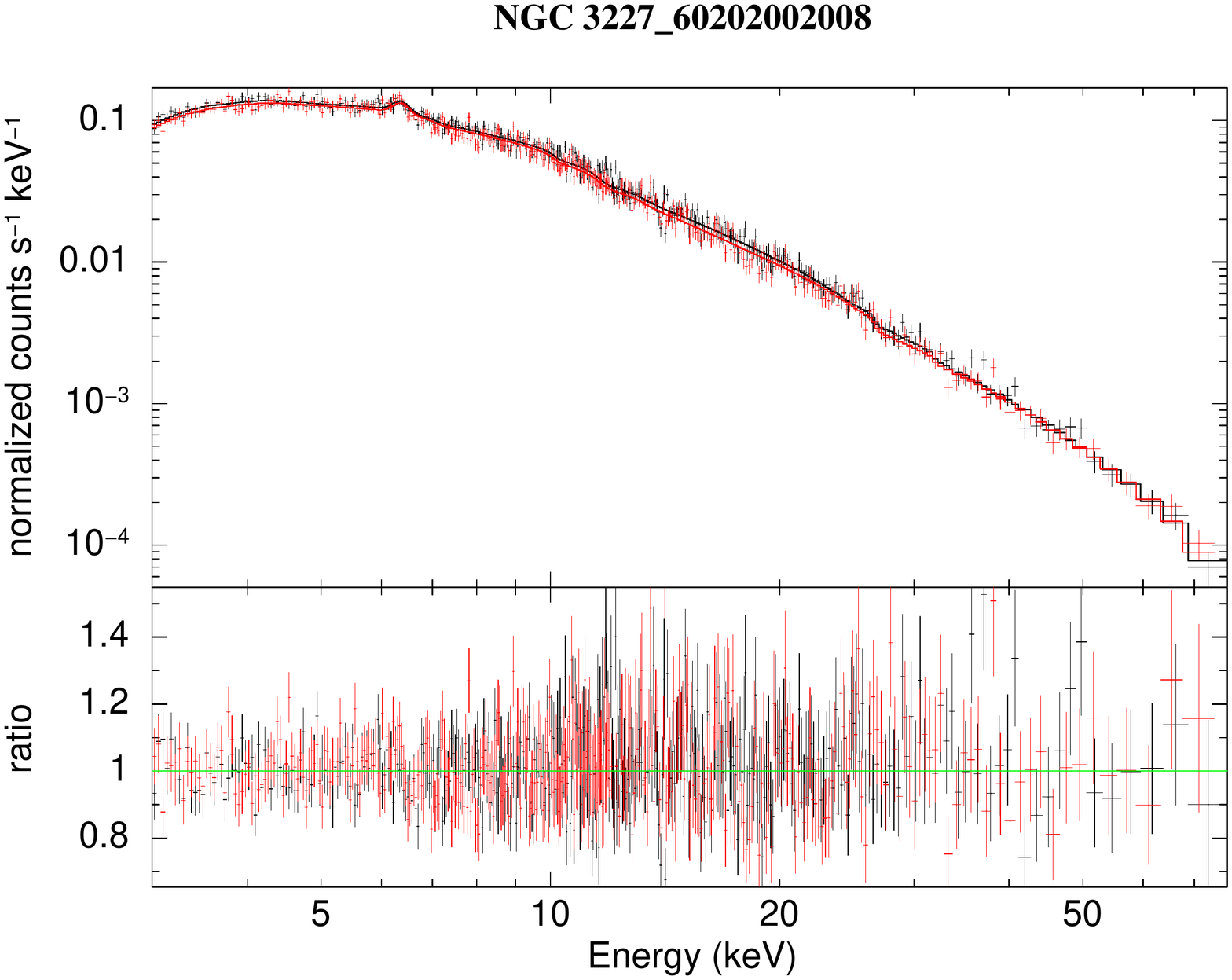}
\includegraphics[width=2.2in]{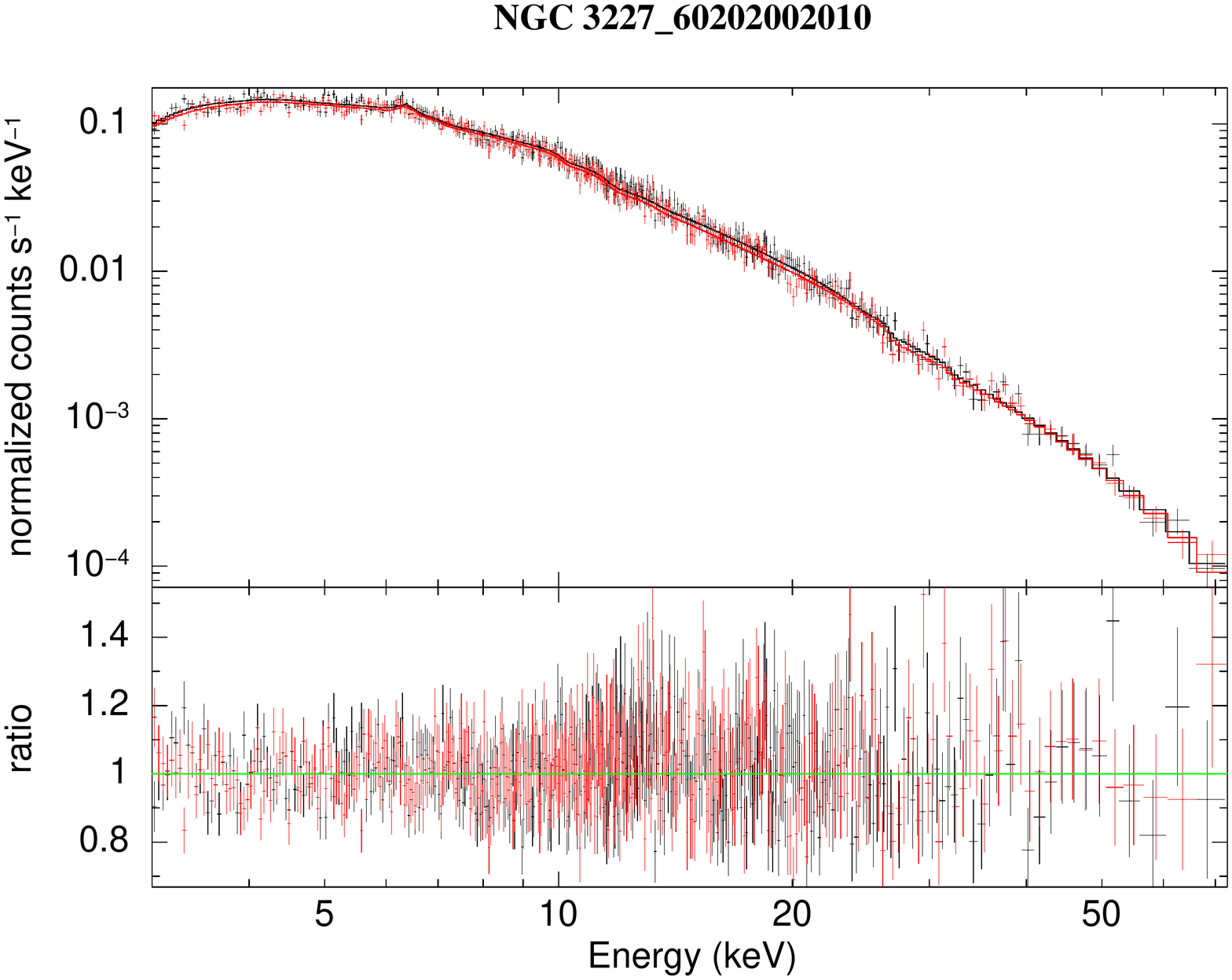}
\includegraphics[width=2.2in]{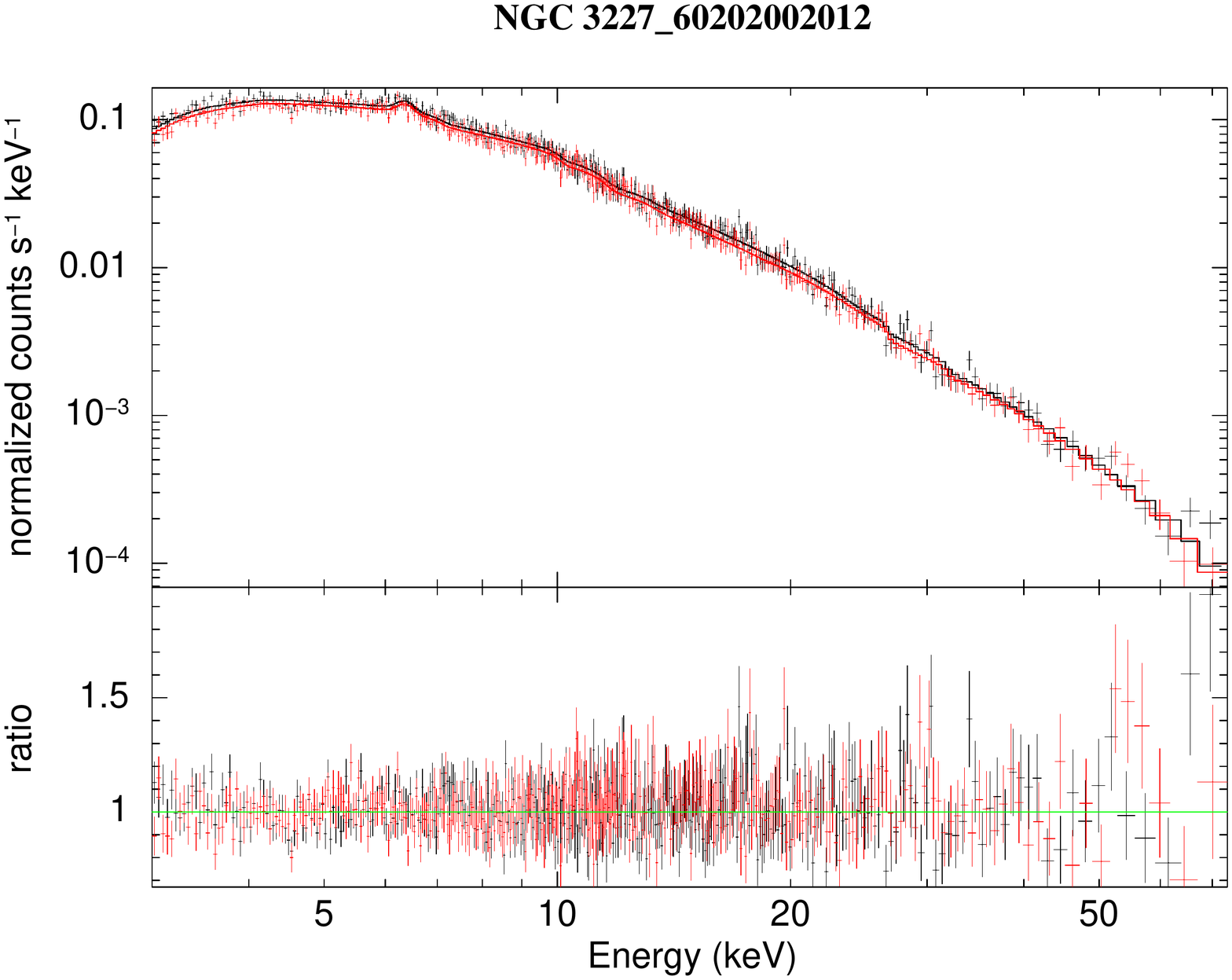}\\
\includegraphics[width=2.2in]{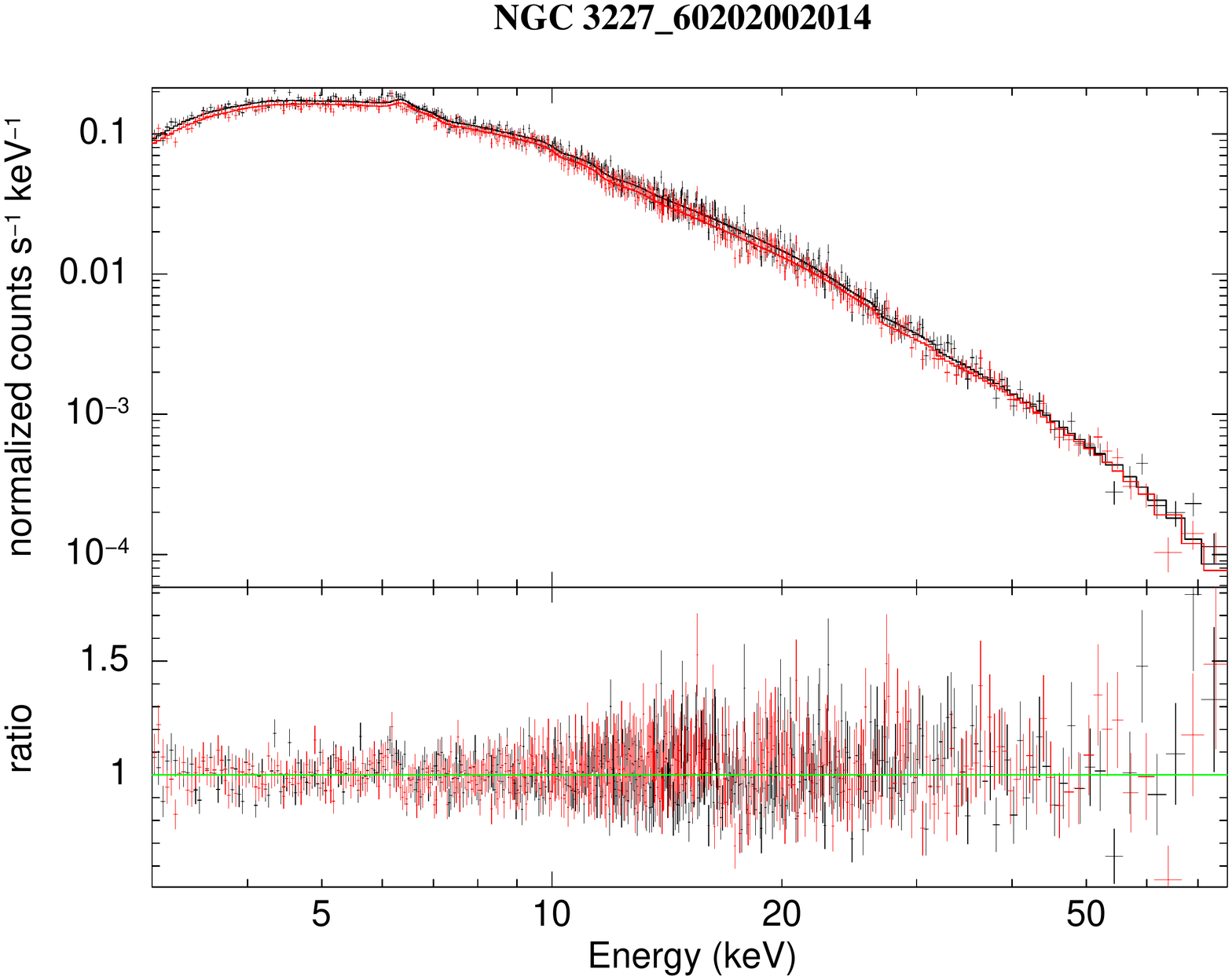}
\includegraphics[width=2.2in]{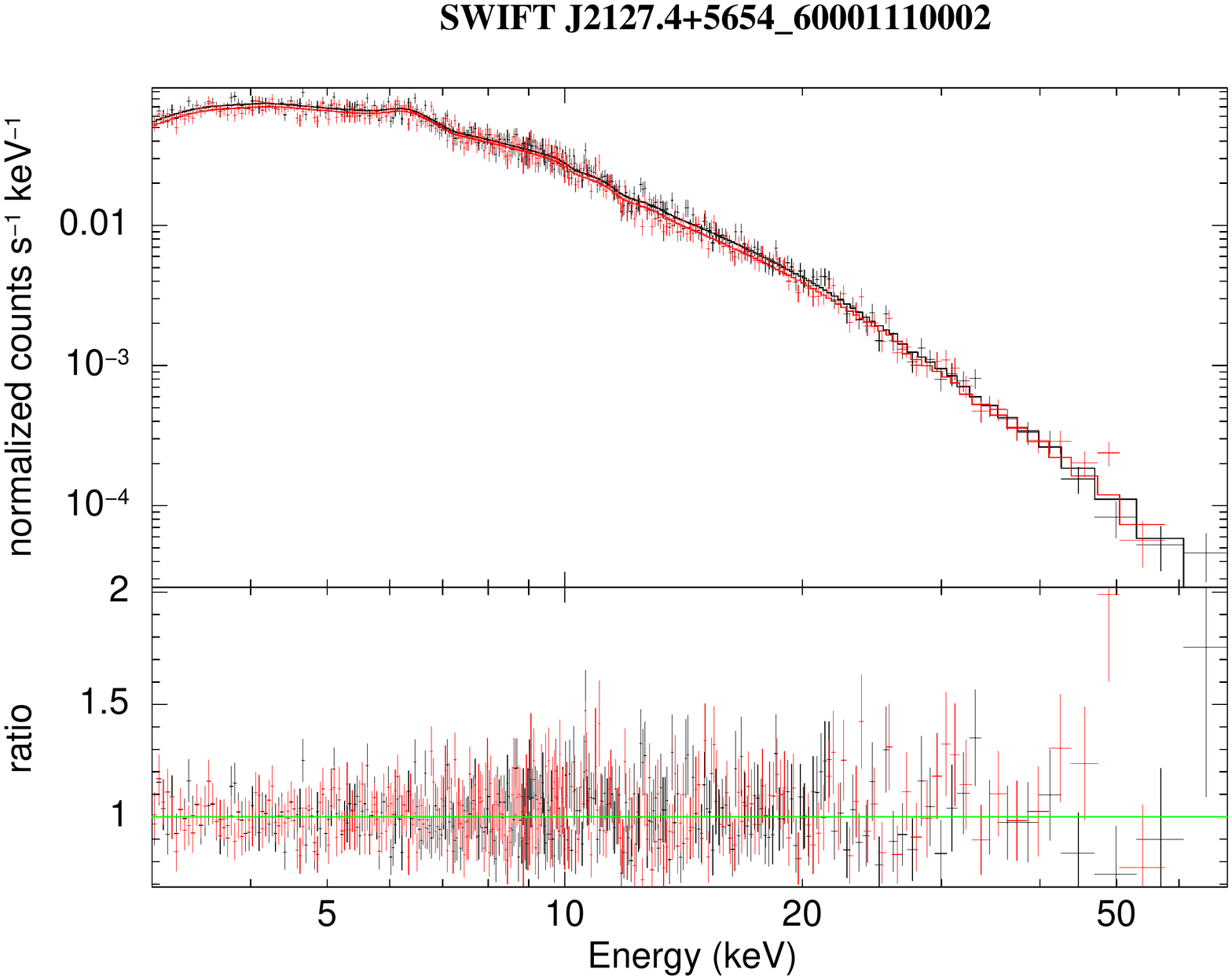}
\includegraphics[width=2.2in]{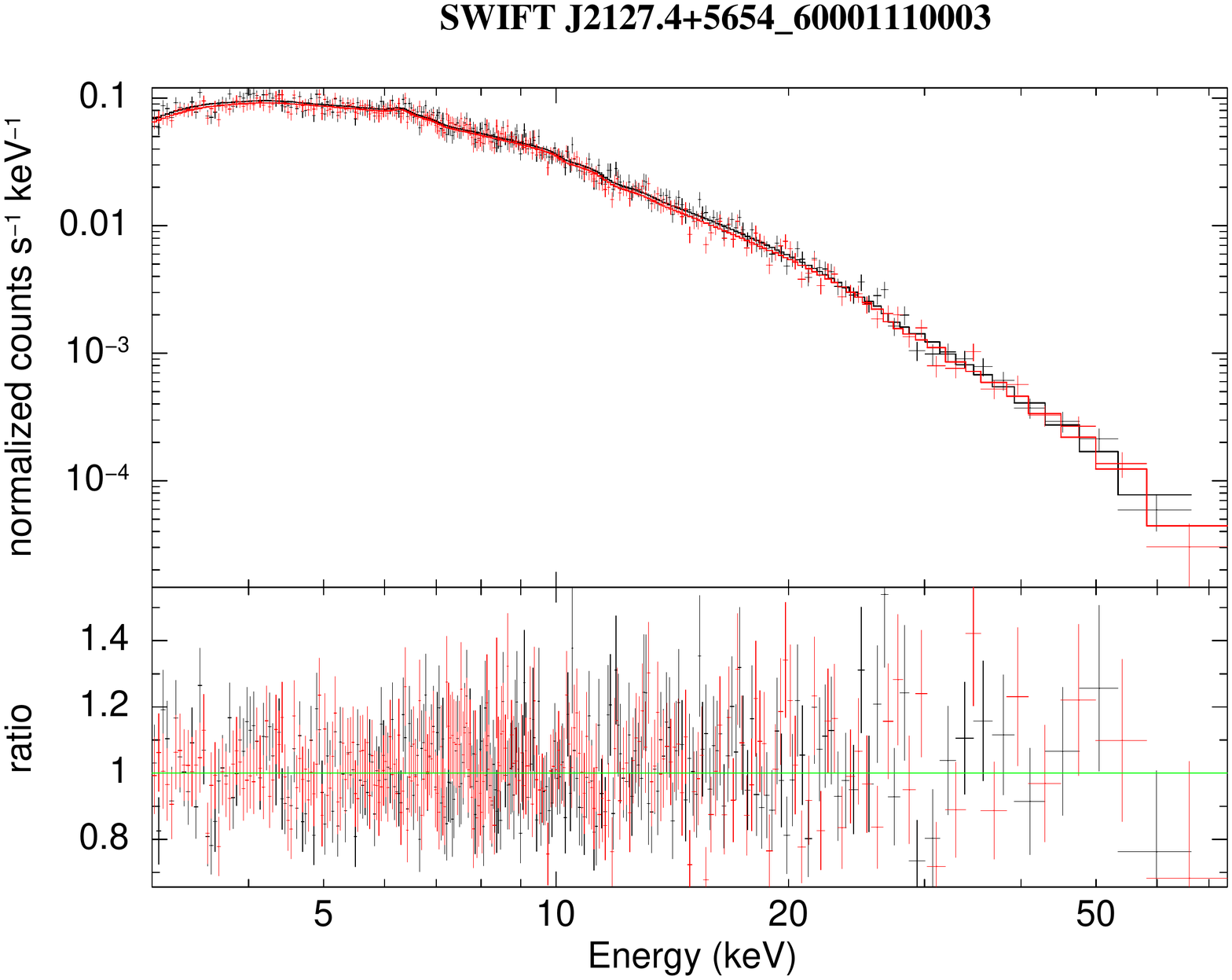}\\
\includegraphics[width=2.2in]{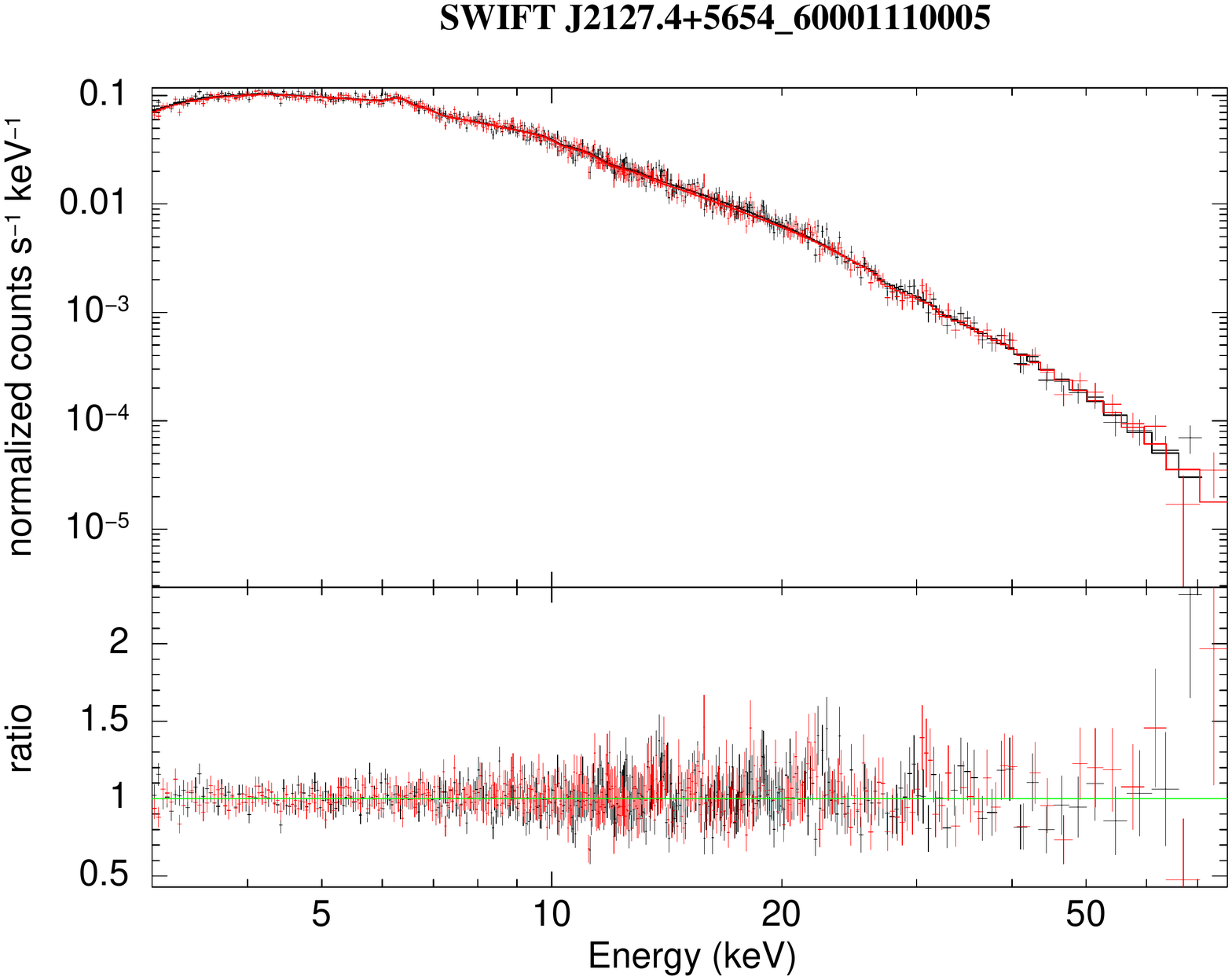}
\includegraphics[width=2.2in]{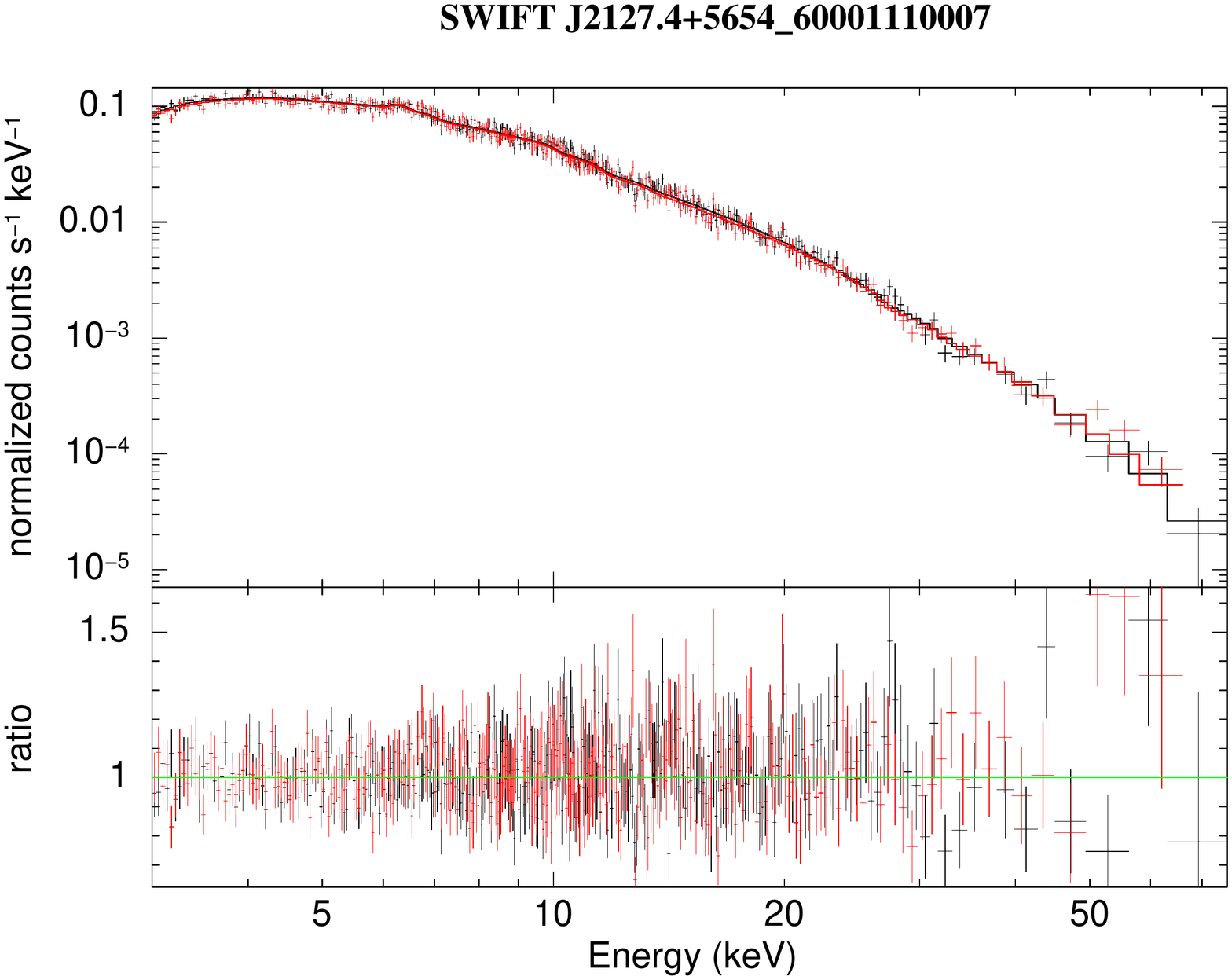}
\includegraphics[width=2.2in]{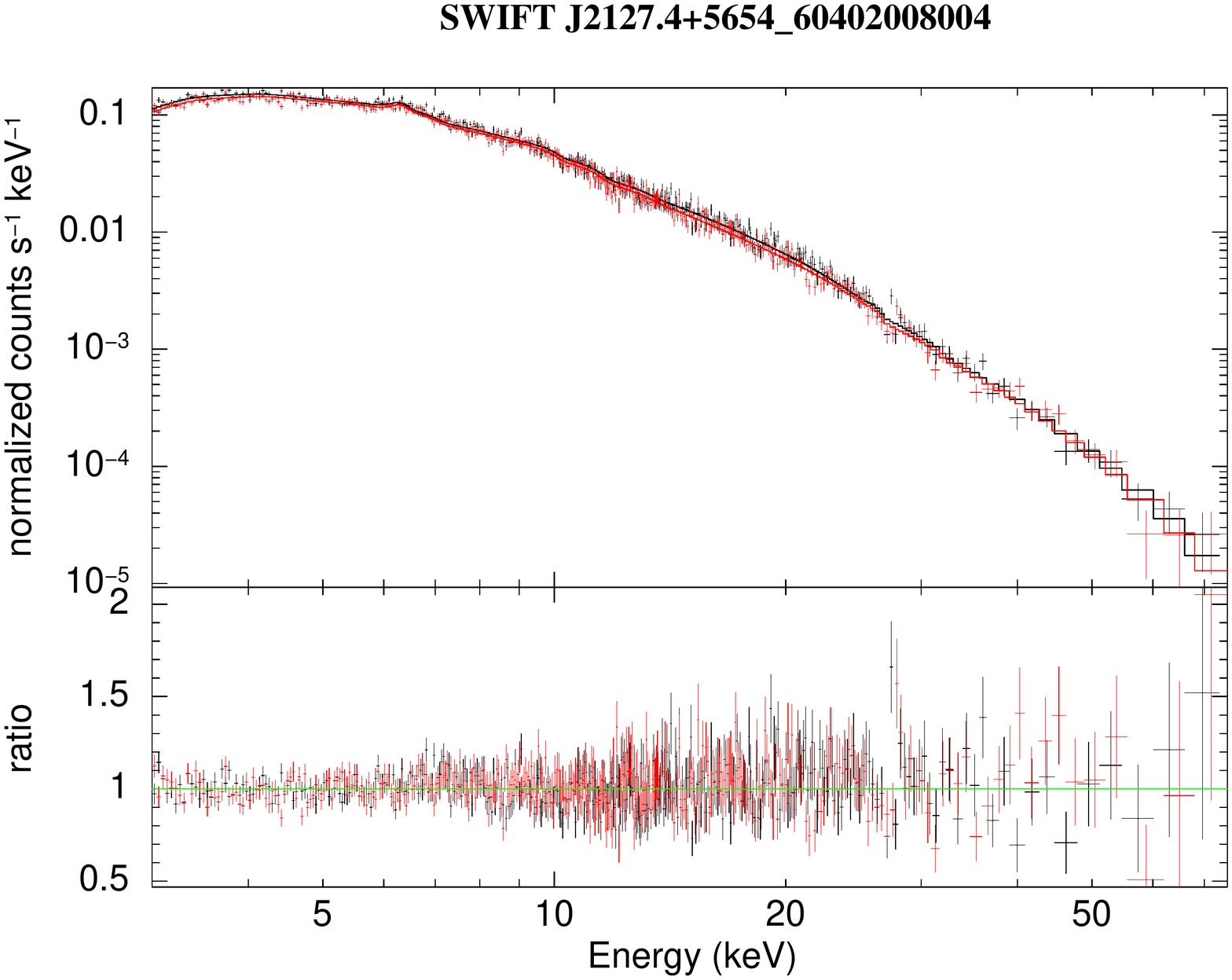}\\
\includegraphics[width=2.2in]{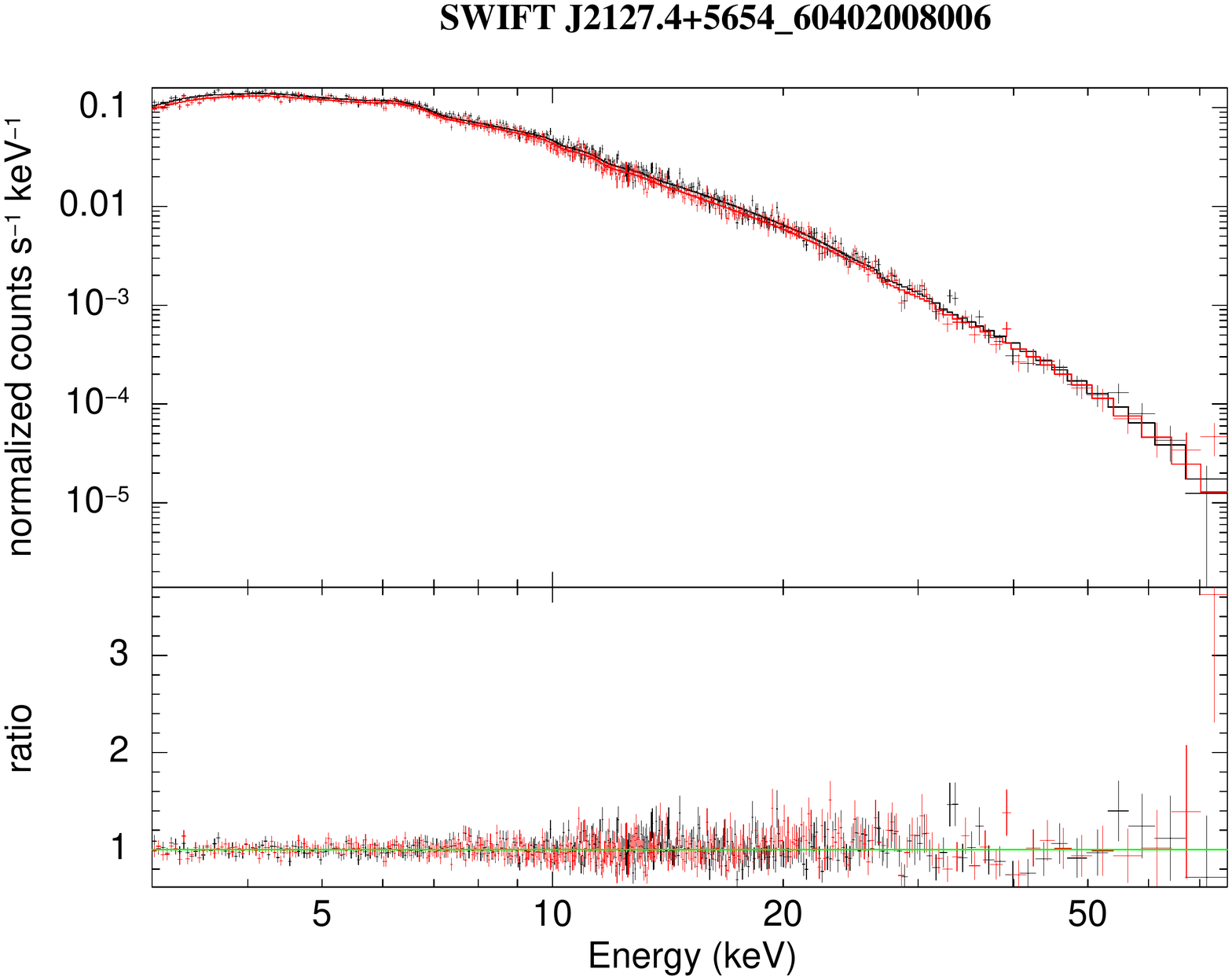}
\includegraphics[width=2.2in]{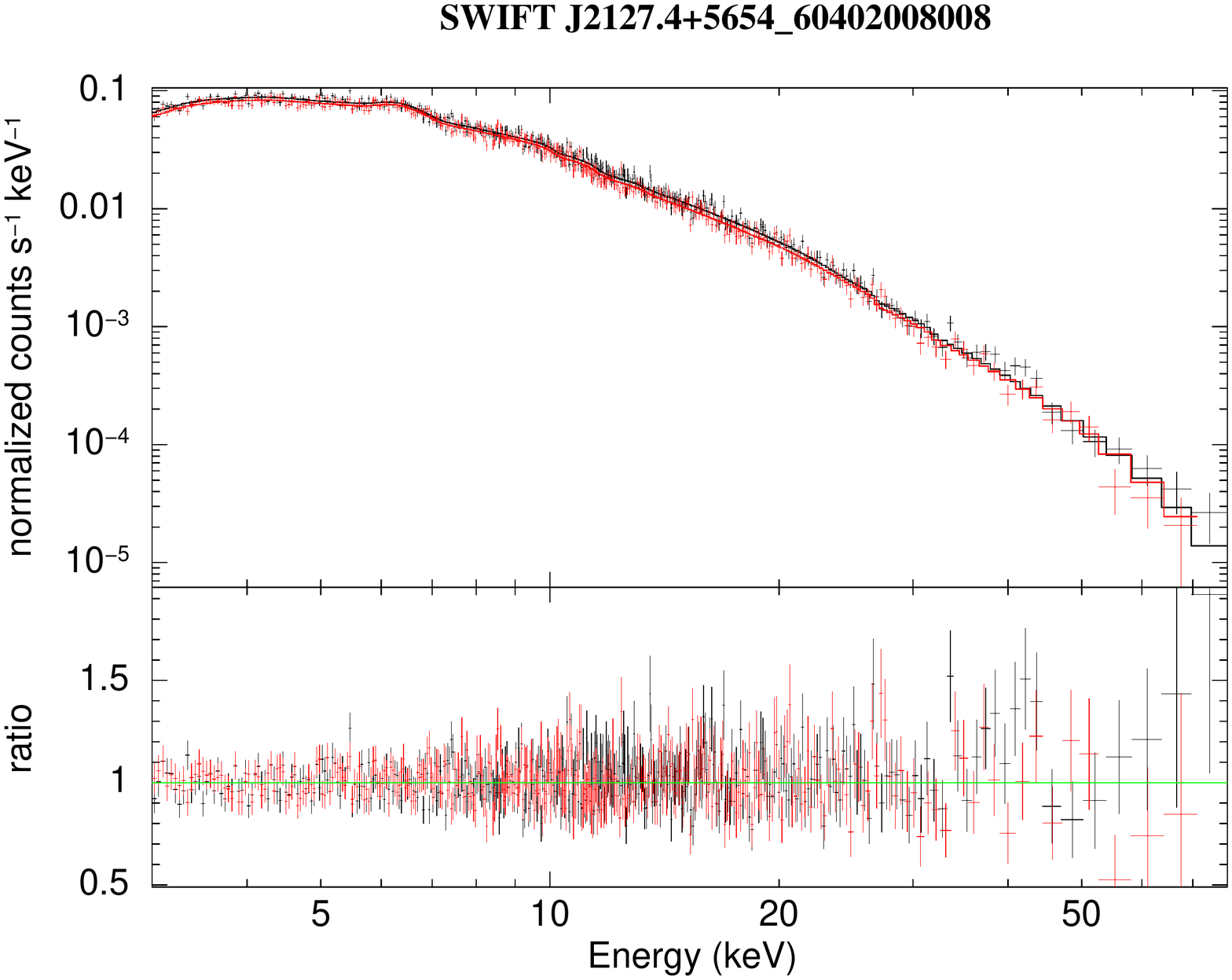}
\includegraphics[width=2.2in]{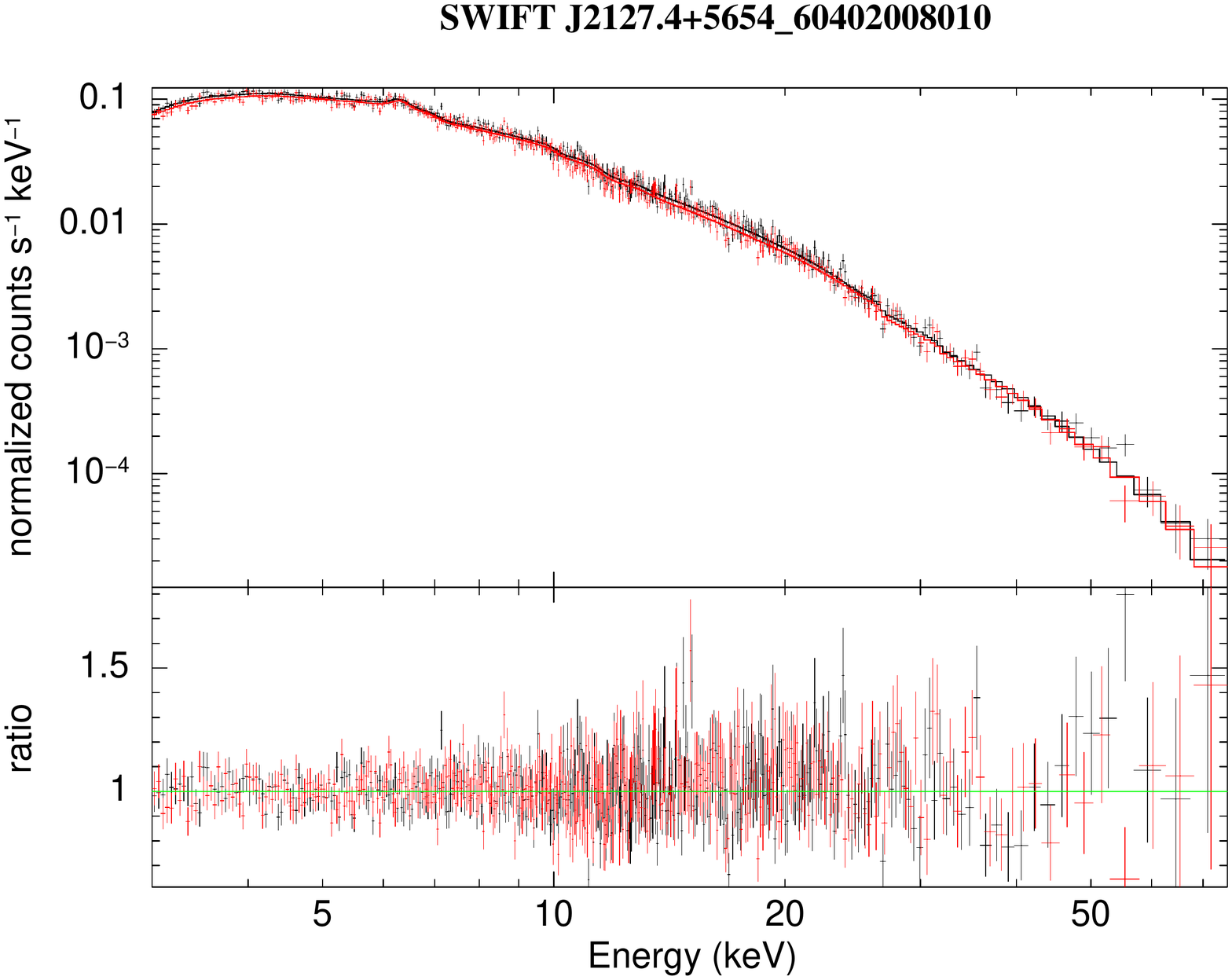}\\
\caption{NuSTAR spectra, best-fit models ({\it pexrav}) and the residual data-to-model ratios. Spectra from both FPMA (black) and FPMB (red) modules are given. The best-fit spectral parameters are given in Tab. \ref{tab:results}. 
 }
\label{fig:all_spectra}
\end{figure*}

\bsp	
\label{lastpage}
\end{document}